\documentclass{aa}

\usepackage{txfonts}
\usepackage{xcolor, colortbl}
\usepackage{graphicx}
\usepackage{array}
 \usepackage{float}
\usepackage{amsmath}
\usepackage{hyperref}
\usepackage{ulem}
\usepackage{makecell}
\usepackage{tikz}
\usepackage{cleveref}
\usepackage{mathtools, nccmath}
\usepackage{multirow}
\usepackage[shortlabels]{enumitem}
\usepackage{pgf}
\hypersetup{
    colorlinks,
    linkcolor={red!50!black},
    citecolor={blue!50!black},
    urlcolor={blue!80!black}
}

\usepackage{silence}
\makeatletter

\renewcommand*\aa@pageof{, page \thepage{} of \pageref*{LastPage}}
\makeatother
\bibpunct{(}{)}{;}{a}{}{,} % to follow the A&A style
\newcommand{\remove}[1]{}
\newcommand{\new}[1]{#1}
\newcommand{\nnew}[1]{#1}

\newcommand{\nremove}[1]{}

\begin{document}

   \title{Combining reference-star and angular differential imaging for high-contrast imaging of extended sources}%to improve recovery of extended signals in high-contrast imaging}
    \titlerunning{Combing reference star library and angular differential imaging algorithms}
    
    \author{S.~Juillard\inst{1}\fnmsep\thanks{F.R.S.-FNRS PhD Research Fellow} 
          \and
          V.~Christiaens\inst{1}\fnmsep\thanks{F.R.S.-FNRS Postdoctoral Fellow}
          \and
          O.~Absil\inst{1}\fnmsep\thanks{F.R.S.-FNRS Senior Research Associate}
          \and
          S.~Stasevic\inst{2,3}
          \and
          J.~Milli\inst{2}}
      
    \institute{STAR Institute, Universit\'e de Li\`ege, All\'ee du Six Ao\^ut 19c, 4000 Li\`ege, Belgium \\
    \email{sjuillard@uliege.be}
    \and
    Univ. Grenoble Alpes, CNRS, IPAG, 38000 Grenoble, France
    \and 
    LESIA, Observatoire de Paris, Universit\'e PSL, Sorbonne, Universit\'e Paris Cit\'e, CNRS, 5 place Jules Janssen, 92195 Meudon, France}
    
    \date{Received 26 February 2024, Accepted 03 June 2024}

% \abstract{}{}{}{}{} 
% 5 {} token are mandatory
 \abstract
  {High-contrast imaging (HCI) is a technique designed to observe faint signals near bright sources, such as exoplanets and circumstellar disks. The primary challenge in revealing the faint circumstellar signal near a star is the presence of quasi-static speckles, which can produce patterns on the science images that are as bright, or even brighter, than the signal of interest. Strategies such as angular differential imaging (ADI) or reference-star differential imaging (RDI) aim to provide a means of removing the quasi-static speckles in post-processing.} 
   {In this paper, we present and discuss the adaptation of state-of-the-art algorithms, initially designed for ADI, to jointly leverage angular and reference-star differential imaging (ARDI) for the direct HCI of circumstellar disks.}%protoplanetary systems.}
   {Using a collection of HCI datasets, we assessed the performance of ARDI in comparison to ADI and RDI based on iterative principal component analysis (IPCA). These diverse datasets were acquired under various observing conditions and include the injection of synthetic disk models at various contrast levels. We also considered reference stars with different levels of correlation with the science targets.}
   {Our results demonstrate that ARDI with IPCA improves the quality of recovered disk images and the sensitivity to planets embedded in disks, compared to ADI or RDI individually. This enhancement is particularly pronounced when dealing with extended sources exhibiting highly ambiguous structures that cannot be accurately retrieved using ADI alone, and when the quality of the reference frames is suboptimal, leading to an underperformance of RDI.  We finally applied our method to a sample of real observations of protoplanetary disks taken in star-hopping mode, and propose to revisit the protoplanetary claims associated with these disks. Among \remove{seven}\new{eight} proposed protoplanets claimed through velocity kinks or direct imaging, none of them were re-detected in our new processed images. }
  {}


   \maketitle

%% ---------------------------- Introduction ------------------------------------------- %%

\section{Introduction}
\label{sec:intro}

   High-contrast imaging (HCI) is a technique that aims to observe faint signals near bright sources, such as exoplanets and circumstellar disks. After the majority of the starlight and aberrations caused by atmospheric turbulence have been removed by the coronagraph and the adaptive optics (AO), respectively, residual atmospheric speckles and quasi-static speckles not seen by the AO system remain \citep{Racine99}. These speckles must be removed to reveal the faint circumstellar signal hiding in the vicinity of the star. Several observing strategies aim to provide a lever to distinguish the speckle field from the disk and planet signal, such as spectral differential imaging \citep[SDI,][]{Racine99, Marois00} and polarimetric differential imaging \citep[PDI,][]{Kuhn01}. In this paper, we focus on \nremove{the following} two strategies. \nnew{Firstly, }angular differential imaging \citep[ADI,][]{Marois06} consists of capturing images over time while targeting a star and fixing the orientation of the telescope pupil, causing the field of view to rotate due to the Earth's rotation. The discriminant employed by the ADI strategy to differentiate the speckle from actual signals is that, unlike the field of view that rotates according to the parallactic angle, the speckle field resulting from imperfections in the instrument remains mostly fixed. Angular diversity is typically leveraged by discarding the static component (which represents the speckle field) to retrieve the rotating circumstellar signal. \nnew{Secondly, }reference-star differential imaging \citep[RDI,][]{Mawet09, Lafreni09, Ruane19} involves using observations of reference star(s) that share similar characteristics with the scientific target of interest, either taken during data acquisition or selected from an archive library. The primary goal is to acquire reference images that exhibit a similar speckle pattern. References are leveraged as a model of the speckle field, which is to be subtracted from the observed data after being appropriately scaled in flux.

    %\item  : The fundamental concept behind PDI involves capturing two orthogonally polarized images of the same object simultaneously. Taking their difference aims to eliminate all non-polarized light components, with the primary contributors being the central star's emission and speckle noise. The resulting signal contains only the polarized light arising from scattered light in the disk.

The choice of an appropriate observing strategy depends on the specific scientific goals, the nature of the target being observed, and the quality of available resources. Each of these high-contrast imaging strategies has its strengths and limitations. ADI is effective at distinguishing faint point-like sources, but it does require a sufficient parallactic angle rotation to function correctly and can cause deformations in extended sources due to rotational invariance \citep{Milli}. \new{These deformations depend on the amount of rotation and the corresponding degree of rotational symmetry, considering the disk morphology \citep{Sandrine23}.} This limitation not only hampers most face-on disks from being captured, but more generally applies to any extended source morphology, leading to challenging-to-correct deformations that can mislead the interpretation of the astronomer. Meanwhile, RDI is efficient for stable speckle patterns and can produce high-fidelity images of disks, but generally requires dedicated observations of an appropriate reference star in very similar observing conditions. \remove{It might not be suitable for}\new{It is less effective in the case of} particularly challenging datasets, such as faint disks, time-varying speckle patterns, or when no well-suited reference star is available, which can be the case for archival ADI datasets. 

To best utilize the capabilities of each strategy, employing a suitable post-processing algorithm is required. Principal component analysis \citep[PCA,][]{Soummer12, Amara12} is widely used and can be applied to both ADI and RDI, but it can suffer from overly aggressive PSF (point spread function) subtraction, resulting in deformations of the circumstellar signal \citep{Milli}.
In the case of ADI, PCA and other PSF-subtraction algorithms such as median subtraction \citep[c-ADI,][]{Marois06} and LOCI \citep[locally optimized combination of images,][]{Lafreni07} work well for point-like sources but generally struggle with extended structures such as disks. Despite many efforts to compensate for these limitations, more recent algorithms such as iterative PCA \citep[IPCA][]{mayo,StapperGinski22} or inverse problem (IP) approaches \citep[MAYONNAISE, REXPACO, and MUSTARD;][]{mayo, REXPACO, Sandrine} are still prone to deformations of extended structures, in particular when these structures include signals that appear static throughout the ADI image sequence. This means that angular diversity is not a suitable discriminator to distinguish this type of signal from the quasi-static speckle field.
Regarding RDI, other algorithms such as data imputation with sequential non-negative matrix factorization \citep[DI-sNMF,][]{Ren18} and the Karhunen-Loève transform with data imputation \citep[DIKL,][]{Ren23} have demonstrated efficiency in optimally extracting circumstellar signals and mitigating the risk of over-subtraction. While RDI can be highly effective, it necessitates that the speckle pattern remain relatively stable over time, and requires sufficiently correlated reference stars to produce high fidelity images.

\begin{table}[!t]
\caption{\nremove{}Notations.}
\label{tab:not}
\begin{tabular}{p{0.2\linewidth} p{0.7\linewidth} }
\hline \hline
Notation                        & Definition                                    \\ \hline 
\multicolumn{2}{c}{Constants}                                                   \\ \hline
$m$                               & Width of an image \new{(i.e., the number of pixels on one side of the square images)}                           \\
$n$                               & Number of frames in the ADI cube              \\
$r$                               & Number of frames in the reference library     \\
$q$                               & Number of dimensions in the PCA low-rank subspace (aka rank) \\ \hline
\multicolumn{2}{c}{Matrices}                                                    \\ \hline
$Y\in \mathbb{R}^{n,m\times m}$       & ADI dataset, i.e., a cube of vectorized images            \\
$L_{r} \in \mathbb{R}^{r,m\times m }$    & Library of $r$ reference stars 
\\
$\Bar{S}\in \mathbb{R}^{n,m\times m}$ & Estimated speckle field\new{s for each image in $Y$}\\
$\Bar{d}\in \mathbb{R}^{+,m\times m}$ & Estimated circumstellar \new{signal (forced to be positive)}
\\
$U \cdot \Sigma \cdot V$&  Singular value decomposition of a matrix 
\\
\hline
\multicolumn{2}{c}{Operators}                                                   \\ \hline
$\Vert \_ \Vert$                       & Absolute value 
\\ 
$\mathcal{H}_{q}(Y)$ & Operator ($\mathbb{R}^{n,m\times m}\mapsto\mathbb{R}^{n,m\times m}$) returning the estimated speckle field\new{s} via PCA using $q$ principal components.
\\ 
\new{$\mathcal{H}_{q}^{L_{r}}(Y)$} & \new{Operator ($\mathbb{R}^{n,m\times m}\mapsto\mathbb{R}^{n,m\times m}$) returning the estimated speckle field\new{s} via PCA using the $q$ principal components of a set of reference $L_{r}$}
\\ 
$\vert \_ \vert_{1}$ & $l_{1}$-norm, sum of the absolute value of all elements in a matrix
\\
$\mathcal{Q}_{Y}$ & Rotation operator ($\mathbb{R}^{m\times m} \mapsto\mathbb{R}^{n, m\times m}$) creating a cube of $n$ images where each frame contains the same image rotated according to the parallactic angles of the sequence of observations $Y$. \new{This operator is used to inject simulated disks or planets into the data.}
\\
$\mathcal{Q}_{Y}^{-1}$ & \new{Median image of a cube after de-rotation} ($\mathbb{R}^{n,m\times m} \mapsto \mathbb{R}^{m\times m}$) \remove{Derotation operator returning the median image of the derotated input.} 
\\
$[\_\;;\;\_]$ &  Concatenation of two collections of images.
\\
\hline
\end{tabular}
\end{table}

Combining diverse observing strategies might assist in mitigating the distinct weaknesses of each approach, as exemplified by \citet{Lawson22}, who combined RDI with PDI, or by \new{\citet{Wahhaj13}}, \citet{Christiaens19}, and \citet{Flasseur22}, who merged ADI with SDI (ASDI) for extended sources. Previous studies in the scientific literature have explored the simultaneous use of RDI and ADI. These attempts typically involve using techniques such as median subtraction or PCA, and applying them to the concatenation of the reference and science cubes \citep[e.g.,][Wallack et al.\ in press]{Carter23}. However, there has not been a dedicated effort to identify the most effective method of combining ADI and RDI, especially in the context of disk imaging. Previous research in this context was rather focused on comparing the performance of ADI with RDI \citep{Ruane19, Xie22}. In response to the described challenges and inspired by the many efforts made in the past few years to develop efficient novel algorithms to retrieve extended sources using ADI and RDI strategies separately, this paper aims to explore how one can adapt the IPCA algorithm to jointly leverage angular and reference-star differential imaging (ARDI) in a simultaneous approach. In Sect.~\ref{sec:Art}, we explain the theoretical foundation and limitations of state-of-the-art IPCA algorithms employing ADI and RDI separately. In Sect.~\ref{sec:ARDI}, we explain how we adapted the IPCA algorithm to jointly leverage ADI and RDI. We also explore the question of parameter optimization. For Sect.~\ref{sec:test}, we conducted a comprehensive comparative evaluation of ARDI against RDI and ADI, all using an IPCA-based algorithm, and assessed their performance across diverse scenarios. In Sect.~\ref{sec:test_real}, we explain how we tested IPCA with ARDI on a sample of real observations of protoplanetary disks, and reevaluated the planet candidates claimed in these disks.

\textit{Notations.} In the rest of the paper, we use the following notations, \new{which are shared with a previous publication \citep{Sandrine23}}: $Y\in \mathbb{R}^{n,m\times m}$ denotes the ADI sequence (also known as the ADI cube), with $n$ being the number of frames, and where each frame is a vectorized square image of size $m\times m$ pixels. The ADI sequence is composed of the circumstellar signal $d$ and the speckle field $S$. The estimate of a parameter is written with an over bar (e.g., $\Bar{S}$). We use $\Bar{S}\in \mathbb{R}^{n,m\times m}$ to denote the cube of the estimated speckle field, which is \new{composed of $n$ unique frames}\remove{unique to each frame}. We use $\Bar{d}\in \mathbb{R}^{+,m\times m}$ to denote the estimation of the circumstellar signal\new{, which is one single positive image, commonly shared by all frames of the cube}. Finally, $L_{r} \in \mathrm{R}^{r,m\times m}$ denotes the library of reference stars containing $r$ frames. Matrix products are written using a dot symbol ($\cdot$). A complete list of all the notation used in this paper is presented in Table~\ref{tab:not}.
    
%% ---------------------------- Section inverse problem ------------------------------------------- %%

\section{IPCA algorithms for ADI and RDI processing}
\label{sec:Art}

PCA is an unsupervised, model-less, statistical procedure that involves creating an orthogonal subspace to describe data through singular value decomposition (SVD). SVD generalizes the eigendecomposition of a square normal matrix with an orthonormal eigenbasis matrix. This decomposition is expressed as $Y = U \cdot \Sigma \cdot V$, where $Y \in \mathbb{R}^{n, m\times m}$ is a vectorized set of images, $\Sigma$  is a diagonal rectangle matrix ($\mathbb{R}^{m, n}$) containing the singular values, and the two matrices $U$ and $V$ represent the left and right singular vectors. In our specific application, $U \in \mathbb{R}^{n, n}$ corresponds to the temporal principal components (PCs) of the data and $V \in \mathbb{R}^{m\times m, m\times m}$ corresponds to the spatial PCs. They are arranged such that the first PCs represent components that explain most of the variance in the data cube (e.g., the more static part, shared by most images), while the higher-rank PCs represent the more unique features of the dataset, such as the \new{detector} noise. To capture the quasi-static speckle field, we will use the orthogonal subspace formed by $V$. It is composed of principal components  $[v_1, \ldots, v_{m\times m}]$, each having the size of one image ($\mathbb{R}^{m\times m}$). By retaining low rank PCs and discarding higher-rank PCs, it is possible to create a subspace $V_q \in \mathbb{R}^{q, m\times m}$ of $q$ images, forming a collection of PCs $[v_1, \ldots, v_q]$ that capture the most spatially static and stable features of the data (i.e., the quasi-static speckles).
Then, projecting the ADI images onto this subspace enables the extraction of the quasi-static speckle image at each frame such that $ \Bar{S} = Y \cdot V_q \cdot V_q^T$. We define the operator $\mathcal{H}_{q}(Y) = \Bar{S}$ to encapsulate all operations described above to estimate the speckle field. Finally, subtracting $\Bar{S}$ from the original data creates a residual cube that hopefully still contains the circumstellar signals. The disk estimate $\Bar{d}$ is obtained from the median frame of the derotated residual cube.

Despite an appropriate choice for the number of PCs $q$ used to build the speckle field estimate $\Bar{S}$, PCA systematically leads to partial subtraction of extended signals \citep{Esposito14, Christiaens19}. The causes of these deformations were described by \citet{Pueyo16} as over-subtraction, which refers to the partial projection of circumstellar signal onto a low-rank subspace, and self-subtraction, which refers to the signal of interest being partially captured in the PCA low-rank subspace. While these effects can be attributed to the algorithms, they are linked to the validity of the underlying model used in post-processing. In particular, it has been observed through diverse algorithms that constraining the positivity of the circumstellar signal greatly prevents the risk of over-subtraction \citep{Ren18, REXPACO, Sandrine23}. However, when using ADI-only for extended sources, the self-subtraction effect is partially caused by the inherent ambiguity of the model: the rotation-invariant flux. The deformations caused by this ambiguity cannot be corrected solely in post-processing if no additional information is provided \citep{Sandrine23}.

Iterative PCA has demonstrated proficiency in preserving disk signal when leveraging angular diversity \citep{mayo, StapperGinski22, Sandrine23}. The iterative aspect enables the usage of a model that constrains the positivity of the estimates to enhance the performance of PCA. In this section we reintroduce the IPCA method using either ADI or RDI. These approaches will enable us to establish, in Sect.~\ref{sec:test}, a consistent comparison with our proposed adaption of IPCA for ARDI, as all three strategies, or combinations of strategies, make use of an IPCA-based algorithm. Alongside the presentation of these algorithms, we propose an explanation of the concept of a fixed point, which is fundamental to IPCA algorithms. Understanding the concept of a fixed point is not only essential for these algorithms, but also crucial for appreciating parameter optimization.

\subsection{ADI with IPCA}
\label{sec:adi-ipca}

IPCA consists of iteratively performing PCA on the science data cube $Y$ while subtracting the previously estimated disk signal $\Bar{d}_{i}$ at each step, while imposing positivity, thereby preventing over-subtraction. \new{
The process starts with $\Bar{d}_0 = 0$ (an image of zeros).} For a given rank $q$, one iterative step is detailed as follows:
\begin{align}
\begin{split}
    \Bar{S}_{i+1} &= \mathcal{H}_{q}(Y - \mathcal{Q}_{Y}(\Bar{d}_i)) \, , \\
    \Bar{d}_{i+1} &= \Vert \mathcal{Q}_{Y}^{-1}(Y-\Bar{S}_{i+1})\Vert  \, ,
    \label{equ:ADIstep}
\end{split}
\end{align}
where $\mathcal{Q}_{Y}: \mathbb{R}^{m\times m} \mapsto \mathbb{R}^{n, m\times m}$ is the rotation operator associated with the data cube $Y$, which creates a cube of $n$ images where each frame contains the same image rotated according to the parallactic angles of the sequence of observations $Y$. The inverse operation $\mathcal{Q}_{Y}^{-1}: \mathbb{R}^{n,m\times m} \mapsto \mathbb{R}^{m\times m}$ corresponds to the median image of the derotated input.
The theoretical foundation of this process hinges on the concept of a fixed-point algorithm. \remove{A fixed point, in mathematical terms, refers to an unchanging point $x$ of a function $f$ that satisfies the equation $f(x_{\rm fix}) = x_{\rm fix}$}\new{A fixed point for a function $f$, is a point that remains unchanged by the function, such that $f(x_{\rm fix}) = x_{\rm fix}$}. The stationary point can be found by iterating over the sequence $x_{i+1} = f(x_{i})$. Optimization methods, such as the well-established Newton's method to find roots or minima, rely on this principle. In our specific application, the aim is to reach the following fix-point $d_{\text{fix}}$, such that $ d_{\text{fix}} = \Vert \mathcal{Q}_{Y}^{-1}(Y - H_{q}(Y - \mathcal{Q}_{Y}(d_{\text{fix}})))\Vert $, while finding the optimal description of the quasi-static component via PCA using $q$ components: $\Bar{S} = \Vert H_{q}(Y - \mathcal{Q}_{Y}(\Bar{d})))\Vert $. The rank $q$ determines the variability of the estimated speckle field. For rank $q=1$, the speckle field is considered to be morphologically static and can only vary in intensity.
Selecting a large value for $q$ incorporates more high-variance components into the speckle field description, providing a more intricate representation, capable of capturing greater variability. However, it is crucial to maintain a low rank to avoid encapsulating \new{astrophysical} circumstellar contributions in the speckle field estimation. This constraint is commonly referred to as the low-rank approximation.

\begin{figure*}[!t]
    \centering
    \includegraphics[width=\linewidth]{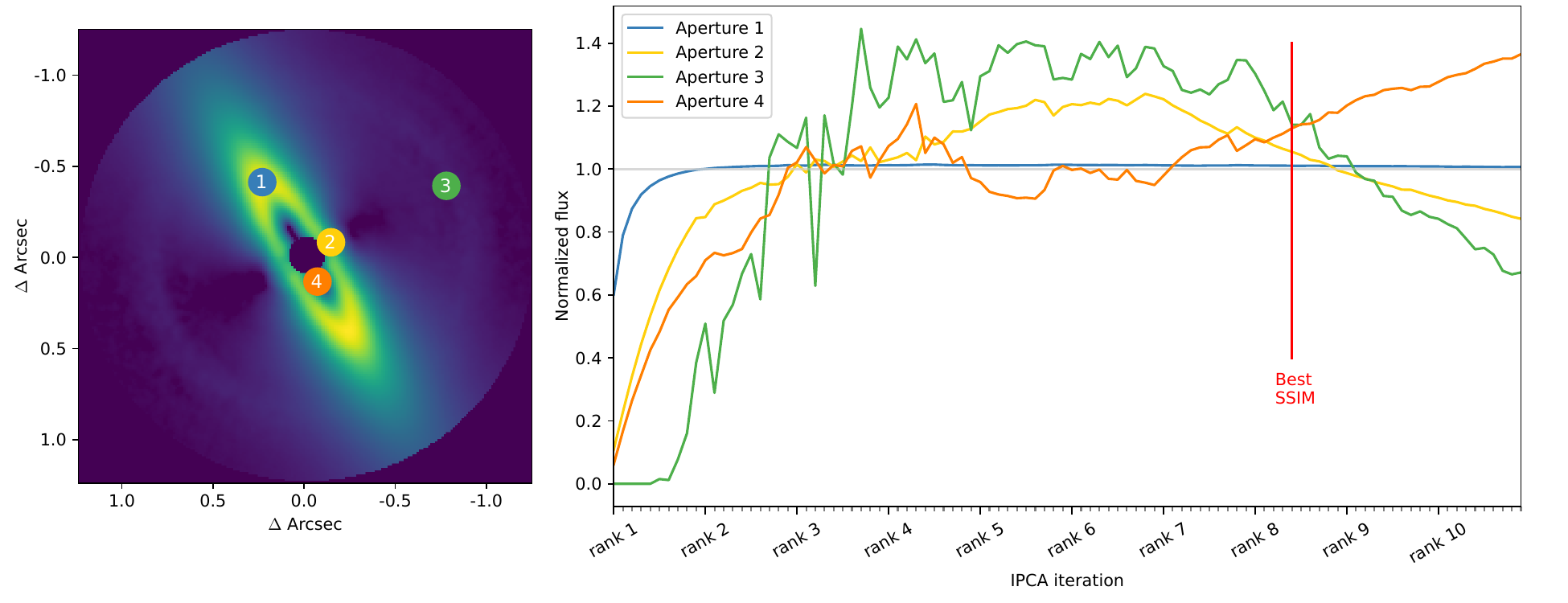}
    \caption{Best disk image, according to the SSIM metric, obtained with the IPCA algorithm leveraging ADI only while processing a synthetic dataset (disk \#B, cube \#3 at a contrast of $10^{-3}$; see Sect. \ref{sec:test}). The image on the left displays different apertures placed on (1) the brightest part of the disk, (2) a region of the disk that is rotation invariant, (3) a noise region, and (4) the region near the coronagraphic region where the variation of the speckle field is higher. On the right-hand side, we plot the evolution of the integrated flux computed in these different apertures through the IPCA iterations. The red vertical line indicates the position of the best image among all iterations according to SSIM. \remove{Flux variations are} \new{Each curve in the figure is} normalized by \remove{their} \new{its} mean. %($\text{flux(app)}/\text{mean(flux/app)}$).
    The x-axis represents each iteration of the IPCA process, starting at rank 1, up to rank 10, with 10 iterations per rank. The gray horizontal line, placed at $y=1$, indicates the position where the flux is equal to its mean.}
    \label{fig:ADIconv}
\end{figure*}
    
When performing an IPCA step, as described in Eq.~\ref{equ:ADIstep}, it is possible to either keep the same rank $q$, or to increase $q$. Subtracting the disk estimation from the original cube before performing the next PCA operation enables more disk signal to be preserved while increasing the rank. However, at each step, the disk estimate will also contain noise and artifacts that will then propagate through the iterations. Hence, wisely choosing the starting rank $q$, when to increase it through the iterations, and when to end the process, is required to obtain an optimal result.
The choice of parameters will influence the amount of noise propagating in the early iterations, and depends on the contrast and variability of the speckle field. When the low-rank approximation is not suitable for capturing enough of the speckle field, as is the case when the amplitude of the variations in the speckle field is larger than the signal intensity, it is necessary to set the starting rank $q$ to a higher value to avoid the propagation of bright artifacts that appear in the first iterations. For the same purpose, the number of iterations per rank should also be limited, especially at a low rank.
Theoretically, multiple sets of parameters should lead to very similar results. In practice, this statement holds true for well-behaved datasets (e.g., bright disks observed under stable conditions), but it can be more sensitive when strong residuals appear in early iterations.
The estimate reached after a few iterations has been observed to correct for over-subtraction and recover the signal with its correct intensity \citep{StapperGinski22, Sandrine23}. However, it is important to note that when using ADI-only for imaging extended signals, the estimate will still suffer from deformation stemming from rotation-invariant components \citep{Sandrine23}. Indeed, algorithms like PCA are designed to remove all signals that appear static and quasi-static. Therefore, without a method to prevent ambiguous rotation-invariant signals from being assigned to the speckle field, the contribution from the disk will inevitably not be \new{fully} preserved.

\begin{figure*}[!t]
    \centering
    \includegraphics[width=\linewidth]{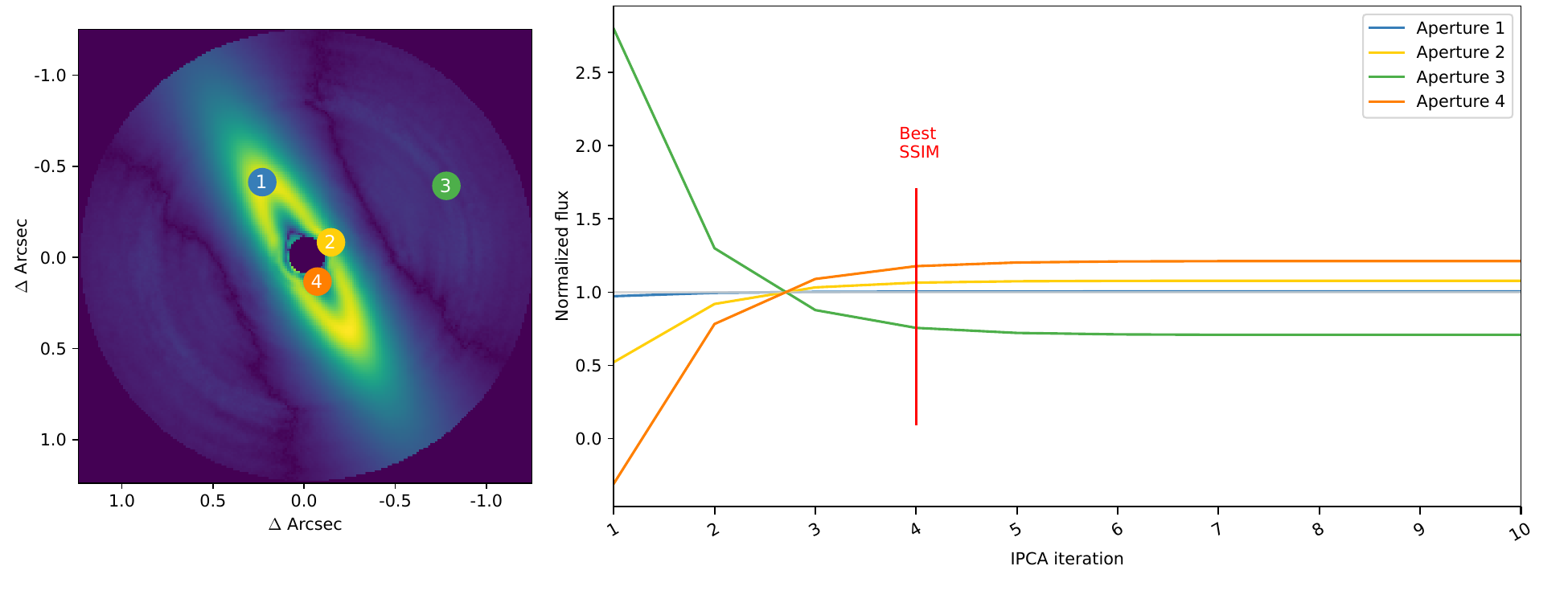}
    \caption{\new{Same as Fig.~\ref{fig:ADIconv}, but for RDI-only IPCA. The set of reference frames used here is presented in Sect.~\ref{sec:test} and is referred to as ``optimal.'' The x-axis represents each iteration of the IPCA process, for ten iterations using a rank-1 PCA.}}
    \label{fig:RDIconv}
\end{figure*}

In Fig.~\ref{fig:ADIconv}, we show the evolution of the disk estimate through IPCA iterations on an example synthetic dataset. The IPCA is parameterized such that it performs ten iterations per rank, while starting at rank $q=1$ and increasing the rank up to $q=10$. The image on the left provides the best estimation of the injected disk according to the structural similarity index measure \citep[SSIM,][]{SSIM}. %This image displays four different apertures placed (1) on the brightest part of the disk, (2) on a region of the disk that is rotation invariant, (3) in the background, and (4) near the coronagraphic region where the variation of the speckle field is larger. On the right-hand side, we plot the variation of the integrated flux computed in these different apertures. 
\new{This metric, which we also use in Sect. \ref{sec:test}, is particularly adapted for assessing image similarity. It evaluates not only the pixel-wise differences between the images but also considers structural information such as contrast, luminance, and structure on various windows of the two images. Indeed, ranking disk image estimations is a multifactorial problem that often necessitates finding a satisfactory compromise between fewer geometrical biases and fewer noise residuals or errors in the overall restored flux.}

We can observe that the flux in the aperture \#1, placed in the brightest part of the disk, converges rapidly and stays stable through the iterations, which is the desired behavior of a reached fixed-point. In contrast, in other areas, such as those with only noise or rotation-invariant flux, no convergence is observed.

% \sj{Il y a aussi le fait que l'opération PCA ne possède pas les protpiétés de continuité et dderivabilité nécessaire pour garentir la convergence. mais aussi le fait que le thorème du point fix ne peut pas garentir que, même si un point fix exsite; est attractif et peut être trouvé.}
% Indeed, at least two major limitations can explain the difficulty to reach convergence over the whole image:

% \begin{itemize}
%     \item Degeneracy of the model: leveraging ADI can be ambiguous due to the variation of the speckle field, but also due to flux invariant to the rotation. 
%     \item Propagation of artifact: at each step, errors from the previous estimate will potentially propagate through the iterative process. It is difficult to properly parameterize the algorithm and to confidently choose between the different results that can be generated using different sets of parameters.
% \end{itemize}

\subsection{RDI with IPCA}

The implementations of the RDI algorithm found in the scientific literature \citep{Schneider09, Soummer12, Ren18} involves the subtraction of an estimated speckle pattern based on a scaled version of the reference images. We propose an adapted version of the IPCA algorithm that employs an iterative process to optimize the projection of the estimated reference. For a given set of references $L_r \in \mathbb{R}^{r, m\times m}$ of $r$ reference images, we subsequently refer to ${V}^{\rm ref}_q \in \mathbb{R}^{q, m\times m}$ as the collection of $q$ PCs estimated by PCA on the reference library. \new{We define the operator $\mathcal{H}_q^{L_r}(Y)$, which, similarly to $\mathcal{H}_q(Y)$, returns $\Bar{S}_{i+1}$ by projecting the science cube into the principal components, which are now computed on the reference library. We note that $\mathcal{H}_q(Y) = \mathcal{H}_{q}^{Y}(Y)$.} 
The subsequent iterative process can then be expressed as:
\begin{align}
\begin{split}
    %\Bar{S}_{i+1} &= \Vert (Y - \mathcal{Q}_{Y}(\Bar{d}_i)) \cdot {{V}^{\rm ref}_{q}}^T \cdot {V}^{\rm ref}_q \Vert  \, , \\
    \Bar{S}_{i+1} &= \mathcal{H}_{q}^{L_{r}}(Y - \mathcal{Q}_{Y}(\Bar{d}_i)) \, , \\
    \Bar{d}_{i+1} &= \Vert \mathcal{Q}_{Y}^{-1}(Y-\Bar{S}_{i+1})\Vert  \, ,
    \label{equ:RDIstep}
\end{split}
\end{align}
where $\Bar{S}_{i+1}$ is obtained by projecting the data cube $Y$ onto the subspace created from the references, while subtracting the previous disk estimation at each step.
Unlike in ADI-based IPCA, the PCs computed via the references are not updated throughout the process. This approach seeks the optimal projection of the reference star PCs on the dataset, while considering disk signal as positive.  

\new{We display in Fig.~\ref{fig:RDIconv} the convergence process to find the optimal projection while using IPCA-RDI of rank $q=1$. This example was produced using the same test dataset as in Fig.~\ref{fig:ADIconv}. In this example, we can observe that the flux exhibits a very clear convergence, which is achieved in only a few iterations.} This \remove{aims}\new{iterative process enables}  the over-subtraction effect to be mitigated\new{, and significantly improves the results compared to classical PCA.} RDI however remains intrinsically sensitive to the stability of the speckle pattern over time, and necessitates the availability of sufficiently correlated reference stars.

Other algorithms, such as DI-sNMF \citep{Ren18}, have also demonstrated efficiency in preserving the disk signal and preventing over-subtraction. However, for consistency in the test pipeline used for performance comparisons in Sect.~\ref{sec:test}, we use the IPCA implementation for RDI, similar to what we implemented for the ADI and ARDI strategies.

%% ---------------------------- Section

\section{Combining angular and reference-star differential imaging (ARDI)}
\label{sec:ARDI}

\nremove{3.1 ARDI with IPCA}

Simultaneously employing both RDI and ADI strategies is expected to help mitigate the individual limitations of each strategy. The combination of RDI with ADI holds the potential of offering several advantages in the field of high-contrast imaging for exoplanet and disk detection around stars.
Here, our primary focus is to adapt the IPCA approach, initially designed for ADI, to leverage ARDI. Other advanced algorithms using both ADI and RDI, including IP-based approaches, could also be considered, but are beyond the scope of this paper. %The selection IPCA is motivated by its demonstrated proficiency in mitigating self-subtraction issues inherent in ADI datasets and the absence of priors \citep{Sandrine23, StapperGinski22}.
In this section we also explore the problem of parameter optimization for IPCA.

% The decision to not present the full results of these algorithms in the comparative tests in Section~\ref{sec:test} stemmed from the fact that, for the purpose of assessing the potential of ARDI compared to RDI and ADI alone, it was more convenient to compare similar algorithms for leveraging them, and IPCA worked perfectly well for this task. Additionally, we did not find it necessary to extend the article by including algorithms for which we did not observe significant improvements and, in some cases, even noted decreased performance, all without achieving improved practicality or resource efficiency. It's important to note that our investigations into the potential implementations of these algorithms are not exhaustive, leaving the door open for possible future improvements.

\label{sec:ARDI-ipca}

\begin{figure*}[!t]
    \centering
    \includegraphics[width=\linewidth]{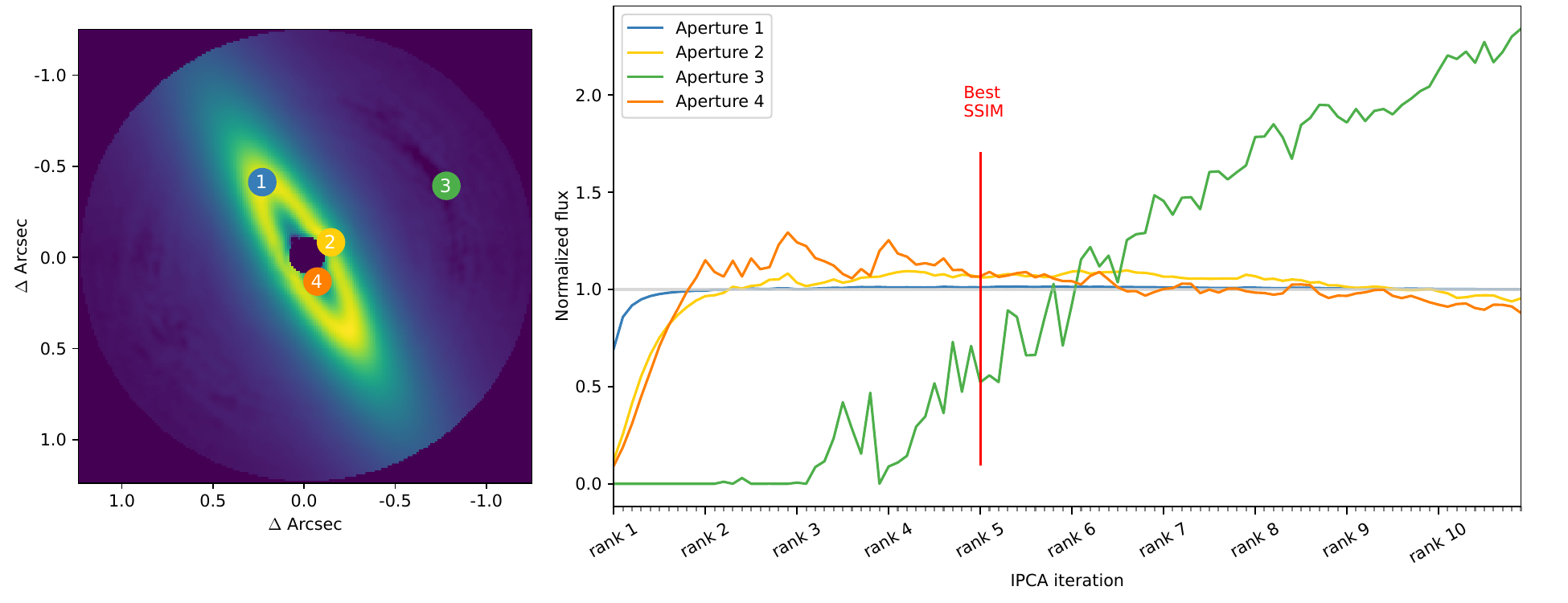}
    \caption{Same as Fig.~\ref{fig:ADIconv} but by leveraging the ARDI strategy with an IPCA algorithm, \new{using the same reference frames as in Fig.~\ref{fig:RDIconv}.} \remove{The set of reference frames used here is presented in Sect.~\ref{sec:test} and is referred to as ``optimal''.}}
    \label{fig:ARDIconv}
\end{figure*}

To combine RDI with ADI using IPCA, we propose injecting the reference star(s) into the ADI cube to compute the PCA low-rank subspace, by concatenating the most correlated reference frames to the ADI cube of images, so that both are used simultaneously to build the speckle field estimate through PCA. The process can be expressed as follows:
% \begin{align*}
%     \Bar{S}_{i+1} &= \Vert H_{q_i}([Y - R(\Bar{d}_n), L_{x}])\Vert  \, , \\
%     \Bar{d}_{i+1} &= \Vert R^{-1}(Y-\Bar{S}_{i+1})\Vert  \, ,
% \end{align*} 
\begin{align}
\begin{split}
     L_{n+r, i} &= [ Y - \mathcal{Q}_{Y}(\Bar{d}_i) \;;\; L_{r}], \\
    \Bar{S}_{i+1} &= \Vert \mathcal{H}_{q}^{L_{n+r, i}}(Y - \mathcal{Q}_{Y}(\Bar{d}_i))\Vert  \, , \\
    \Bar{d}_{i+1} &= \Vert \mathcal{Q}_{Y}^{-1}(Y-\Bar{S}_{i+1})\Vert  \, ,
    \label{equ:ARDIstep}
\end{split}
\end{align}
where the operation $[\_\;;\;\_]$ represents the concatenation of two collections of images, creating a set of images \new{($L_{n+r, i}$)} with a size of $\mathbb{R}^{n+r, m\times m}$ that contains both the data and reference images. The ratio between the size $r$ of the reference library and the number $n$ of frames in the ADI cube must be chosen carefully for optimal results. We refer to $r/n$ as the reference frame ratio. The proposed algorithm can be classified as semi-supervised, as it simultaneously leverages reference frames for the speckle field (which can be regarded as labeled data), and the actual (unlabeled) data to provide an optimal estimation.

\begin{figure}[!ht]
    \centering
    \includegraphics[width=\linewidth]{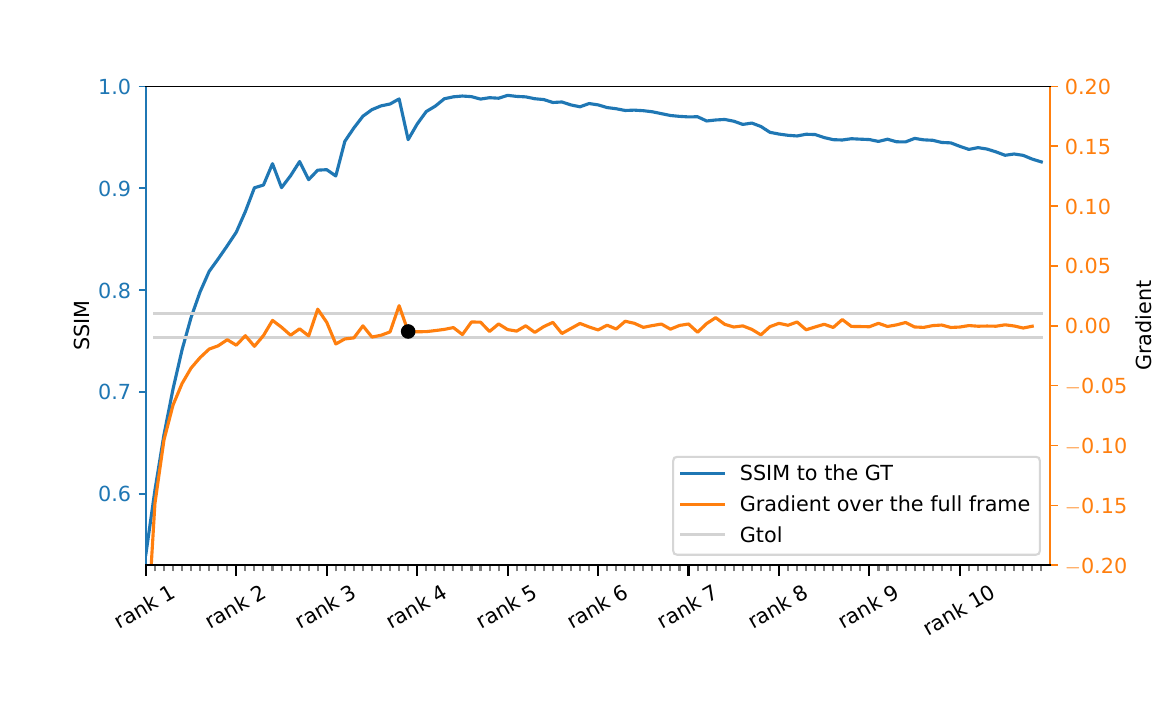}
    \caption{Evolution of the IPCA with ARDI estimate through the iterations, using the same dataset as in Fig.~\ref{fig:ARDIconv}. The blue curve relates to the left y-axis and displays the value of the SSIM for each estimate. The orange curve relates to the right y-axis and represents the gradient computed over the full frame, as described in Eq.~\ref{equ:gtol}. The $g_{\rm tol}$ has been set to $10^{-2}$ and is represented on the figure by the two gray lines. 
    The black dot indicates the iteration where the gradient curve reaches the $g_{\rm tol}$ and does not exceed it in further steps. The x-axis represents each iteration, where the minor ticks represent a single iteration, and the major ticks represent a rank update.}
    \label{fig:fullgradient}
\end{figure}

In Fig.~\ref{fig:ARDIconv}, we monitor the evolution of the disk estimate through the IPCA iterations while using the ARDI strategy, similarly to the test presented in Fig.~\ref{fig:ADIconv} with ADI only. We used identical parametrization (ten ranks, ten iterations per rank), and a reference frame ratio ($r/n$) of one. Conversely to what we observed with ADI only, three out of the four apertures seem to converge when using reference images. Results obtained with ARDI do not exhibit the deformation due to rotation invariant flux (located in aperture \#2). However, the background noise in the final image increases throughout the iterations. 
Consequently, determining where to stop the iterative process involves choosing a good compromise between noise subtraction (non-propagation), filtering, and recovery of extended signals (convergence). A common criterion to assess convergence is setting a stopping criterion based on the gradient of the iterate. In our case, the criterion, based on the disk estimate, reads as follows: 
\begin{align}
     \left\Vert \frac{\vert\Bar{d}_{i}\vert_{1} - \vert\Bar{d}_{i+1}\vert_{1}}{\vert\Bar{d}_{i}\vert_{1}}\right\Vert < g_{\rm tol} \, ,
    \label{equ:gtol}
\end{align}
where $g_{\rm tol}$ stands for the gradient tolerance and defines the precision threshold to be reached to assess convergence. It is a threshold to the gradient of the iterate, computed in our case by comparing disk estimates between two consecutive iterations, $i$ and $i+1$. 

We show in Fig.~\ref{fig:fullgradient} the evolution of the SSIM with respect to the ground truth (the injected disk), as well as of the gradient (Eq.~\ref{equ:gtol}), as a function of the iterations. Additionally, we present in Appendix~\ref{ann:patches_gradient} the evolution of the gradient across patches of $15\times15$ pixels. Firstly, we can observe that multiple iterations achieve a similar SSIM, corresponding to different estimates of equal quality.
Secondly, the iteration where convergence is reached according to the gradient criterion occurs close to the iteration where the SSIM is the highest (even though the criterion serendipitously falls into a local minimum in the SSIM curve). \remove{When computing the gradient over the full frames, parameters such as the chosen size of the images or the intensity of the signal compared to the variability of the residual will greatly influence the aspect of the gradient curve, as the flux is integrated over the whole image. This leads to pixels of higher intensity driving the shape of the curve, potentially missing the convergence of the full disk, including its fainter component.} \new{The region over which the gradient is computed can introduce bias, as it may not accurately represent the convergence behavior of the signal of interest. Indeed, if the field of view is too large or if the signal-to-noise ratio \new{of the circumstellar signals} is low, then the regions where there is no signal\remove{will sum up, potentially becoming} \new{can become} dominant in the computation of the gradient. Conversely, a few pixels of significantly higher intensity may dominate the gradient computation, potentially masking the convergence of the full disk, including its fainter components.}
When computing the convergence criterion over patches \new{(Fig.~\ref{fig:patches_gradient})}, we observe that only patches containing signal achieve convergence, and that the brightest part of the disk converges faster than the fainter parts. Overall, we conclude that monitoring the convergence curves can provide a relevant indicator of the optimal iteration to be chosen. Nevertheless, assessing the convergence of a numerical method in computer science is a recurring problem for which proposed solutions are known to lack reliability in some cases \citep{Nocedal2002}. Moreover, it is particularly challenging in our case of application, given that the iterative process is not guaranteed to converge.

\begin{figure*}
    \centering
    \includegraphics[width=\textwidth]{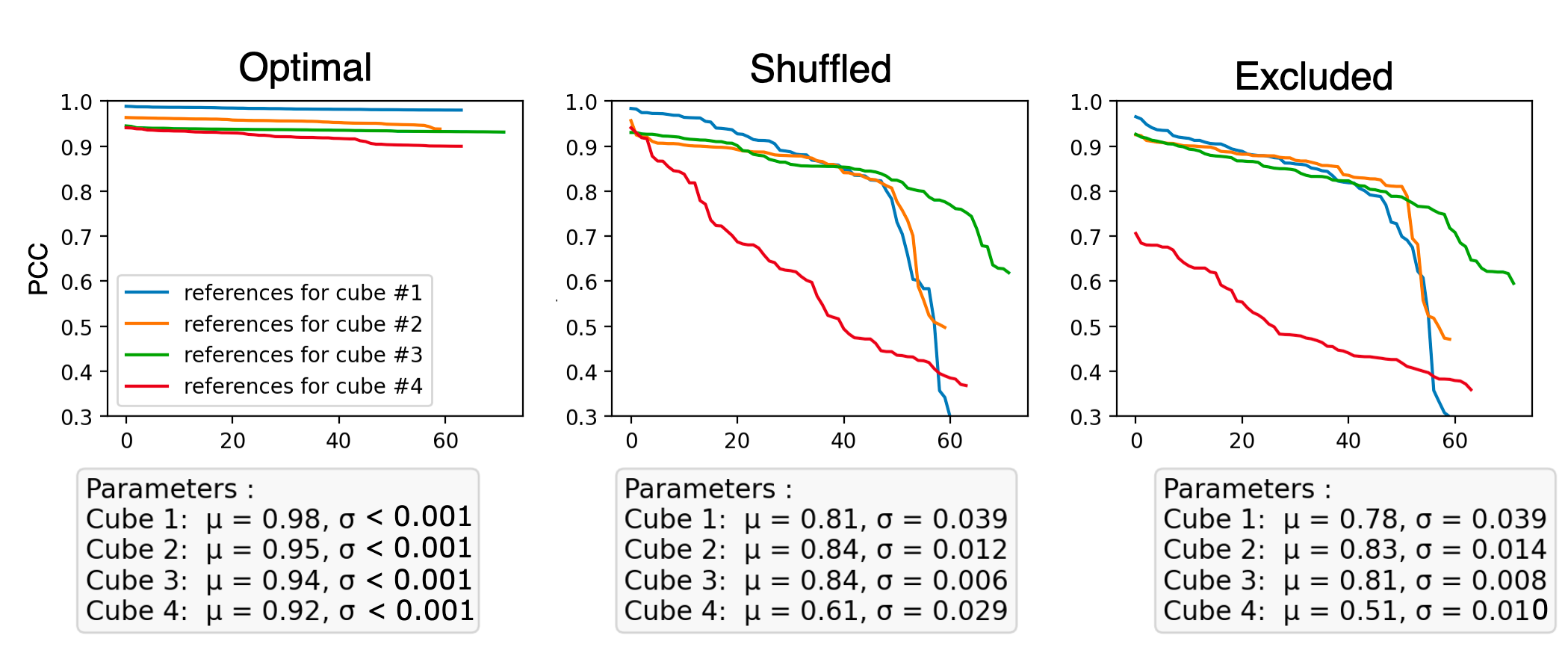}
    \caption{Pearson correlation coefficient of the selected reference frames computed individually and averaged over each frame of the \new{raw} ADI data before the injection \new{of simulated disks or planets}, for three different selections of references: (left) optimal references selected; (middle) a random sample of each cube's dedicated references, including the optimal references, in an even proportion; and (right) a random sample of references excluding the \new{optimal} references. These coefficients were calculated within an annulus with an outer radius of 60 pixels and inner radius of 30 pixels, and are classified from the most correlated to the least correlated. The text box below each figure displays the mean PCC ($\mu$) and the variance ($\sigma$) of the reference sample relative to its associated data cube.}
    \label{fig:pearson_correlation}
\end{figure*}

An alternative method for combining RDI and ADI is to utilize the PCs computed via RDI as the initialization of IPCA with the ADI cube. Hence, only the initialization differs and is expressed by the following equations: 
\begin{align}
\begin{split}
    \Bar{S}_{0} &=  \mathcal{H}_{q}^{L_{r}}(Y) \, , \\
    \Bar{d}_{0} &= \Vert \mathcal{Q}_{Y}^{-1}(Y-\Bar{S}_{0})\Vert  \, .
    \label{equ:RDIinit}
\end{split}
\end{align}
Then, the iterative steps are identical to the IPCA with ADI presented in Sect.~\ref{sec:adi-ipca}. While we did not observe any significant nor systematic improvement when choosing this method, it remains relevant and can in some cases produce results consistent with those obtained with IPCA-ARDI.

%% ---------------------------- Section  ------------------------------------------- %%

\section{Testing algorithms}
\label{sec:test}
\subsection{Tests based on simulated data}
\label{sec:testsim}

In this section we compare the three different strategies: RDI, ADI, and ARDI, all utilizing the same core algorithm, IPCA. We consistently employed the same evaluation pipeline as presented in \citet{Sandrine23}, which was previously used to compare three different algorithms for processing datasets using ADI alone. This test pipeline consists of a total of 60 test datasets composed of five different disk morphologies, injected at three different contrast levels ($10^{-3}$, $10^{-4}$, and $10^{-5}$), into four different observing \new{ADI} sequences of stars without any known circumstellar signal, reflecting different observing conditions. \new{The test datasets are available on Zenodo.\footnote{\nnew{Datasets available at }\url{https://zenodo.org/records/11442267}}}

The datasets, obtained through the High-Contrast Data Center (HCDC), were acquired using the Infrared Dual-Band Imager and Spectrograph \citep[IRDIS,][]{Dohlen08, Vigan10} camera of the Spectro-Polarimetric High-contrast Exoplanet Research coronagraphic system on the Very Large Telescope \citep[VLT/SPHERE,][]{Beuzit19}. The test datasets all consist of the $H$2 channel from the dual-band $H$23 set. They were chosen to exhibit a diverse range of characteristics, including low Strehl ratio with a 26\degr\ rotation (ID\#1), wind-driven halo (ID\#2), an unstable speckle field (ID\#3), and good Strehl ratio with an 80\degr\ field rotation (ID\#4). The raw data processed with the data handling software \citep{Pavlov08} of the HCDC \citep{Delorme17}, which performs dark, flat, and bad pixel correction on a coronagraphic sequence. 
\new{For future reference, we computed the mean and standard deviation of the Pearson correlation coefficients (PCC) between each unique pair of frames in the ADI cube. The mean PCC are as follows: Cube \#1: $\mu = 0.99$; Cube \#2: $\mu = 0.97$; Cube \#3: $\mu = 0.93$; Cube \#4: $\mu = 0.96$, with standard deviations below $0.001$ for all the cubes.}

The injected disks represent a range of scenarios for both debris and protoplanetary disks. \new{As detailed in \citet{Sandrine23}}, this selection consists of two 75\degr\ inclined disks with varying sharpness levels (A and B), a 45\degr\ inclined disk with two concentric rings (C), a nearly face-on disk with azimuthal flux variation (D), and a hydrodynamical simulation of a disk with embedded spiral structures and a companion (E). 
The contrast of the injected disks is determined by measuring the integrated flux within a full width at half-maximum (FWHM)-sized aperture, centered at the peak intensity of the disk, and then dividing this value by the integrated flux within an FWHM-sized aperture of the stellar point spread function. However, we made an exception for the synthetic disk \#E, where we measured the flux at the companion location.

We performed two series of tests. In the first series, we compared the performance of ADI, RDI and ARDI using an IPCA-based algorithm on our 60 test datasets, while using an optimal choice of reference stars. The reference frames were selected from a set of archival IRDIS observations taken with the same filter, coronagraph, and exposure time as the test datasets. These reference targets were observed between 2014 December 11 and 2021 June 1, and the raw data were calibrated through the same process as the test datasets. For each test dataset, the \remove{Pearson correlation coefficient (}PCC\remove{)} was calculated between the frames of the dataset and the reference targets, excluding any observations of the dataset star taken at different epochs. The PCC was calculated within a circular annulus between 0\farcs31 and 0\farcs67, which captures both the dominant speckle region and position of the waffle pattern, used for precise star centering of a coronagraphic sequence \citep{Zurlo14}, if it was included in the observation. For each frame in the dataset, the 300 best correlated reference frames were identified, and those that appeared in this selection for more than 30\% of the dataset frames were selected for the final reference library. 

The left-hand side plot in Fig.~\ref{fig:pearson_correlation} (``Optimal'') displays the PCC for each frame of the reference library. The PCC value for each reference frame corresponds to the average of the correlation coefficients computed individually with each frame of the ADI cube before disk injection. The reference frames are then sorted from the most correlated to the least correlated following this methodology. 

\begin{figure*}[!t]
    \centering
         \includegraphics[width=\textwidth]{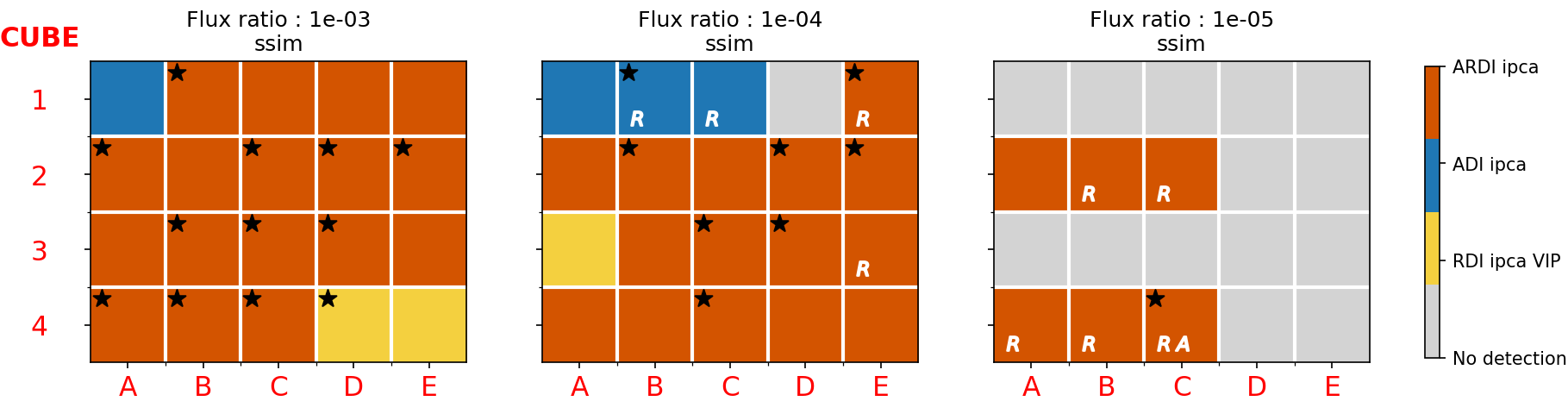}
        \caption{Results of our systematic tests comparing the three strategies RDI, ADI, and ARDI using the IPCA-based algorithm. Each cell represents a different synthetic dataset, combining disk morphologies labeled along the x-axis with datasets before injection along the y-axis. Each figure contains three tables representing three different levels of contrast: $10^{-3}$ (left), $10^{-4}$ (middle), and $10^{-5}$ (right). The color of the cell indicates which algorithm performed best according to the SSIM metric. The white letters on a cell indicate which algorithm(s) did not detect the disk (R\,$\longrightarrow$\,IPCA-RDI, A\,$\longrightarrow$\,IPCA-ADI, $\text{A}_\text{R}\longrightarrow$\,IPCA-ARDI). Gray cells mean that no algorithm detected the disk. Black stars indicate that the five metrics (SSIM, Spearman, Pearson, Euclidean, and SAD) selected the same winner.}
        \label{fig:results}

\end{figure*}

For the second series of tests, we aimed to assess how ARDI could assist in cases of inadequate reference stars. We tested the ARDI-IPCA algorithm for three different qualities of reference frames. In the ``Optimal'' case, we used the most correlated frames, which is the same selection used in the first series, where we compare ADI, RDI, and ARDI. In the ``Shuffled'' case, we randomly selected a sample from the reference libraries dedicated to each of our four ADI \remove{test cubes}\new{sequences} into one common reference library. In this test, the random selection contains 25\% of frames from the ``Optimal'' reference library. In the ``Excluded'' case, we selected for each ADI cube a sample only from references dedicated to the three other ADI cubes, creating a selection that excludes optimal references. The PCC computed for each of the three selections of references is presented in Fig.~\ref{fig:pearson_correlation}.
For the IPCA processing in ADI, RDI, and ARDI, we used our implementation of the \texttt{GreeDS} algorithm\footnote{\url{https://github.com/Sand-jrd/GreeDS}} \citep{mayo, Sandrine23}. This Python implementation utilizes PyTorch, enabling the processing of a dataset from the test pipeline in less than a minute on a standard laptop. For all tests, we used an identical parameterization: ten ranks and a reference frame ratio ($r/n$) of one. The number of iterations increases by one at each rank (e.g., one iteration is performed at the first starting rank $q$, two at rank $q+1$, and so on). This enables improved results compared to setting the number of iterations per rank to 10, as proposed in \citet{Sandrine23}, especially for faint disks. % For the IPCA used in RDI mode, we used the implementation of the Vortex Image Processing package\footnote{\url{https://github.com/vortex-exoplanet/VIP}} \citep{Gomez17, Christiaens23}, with the default parameters.

\begin{figure*}[!t]
        \centering
         \includegraphics[width=\textwidth]{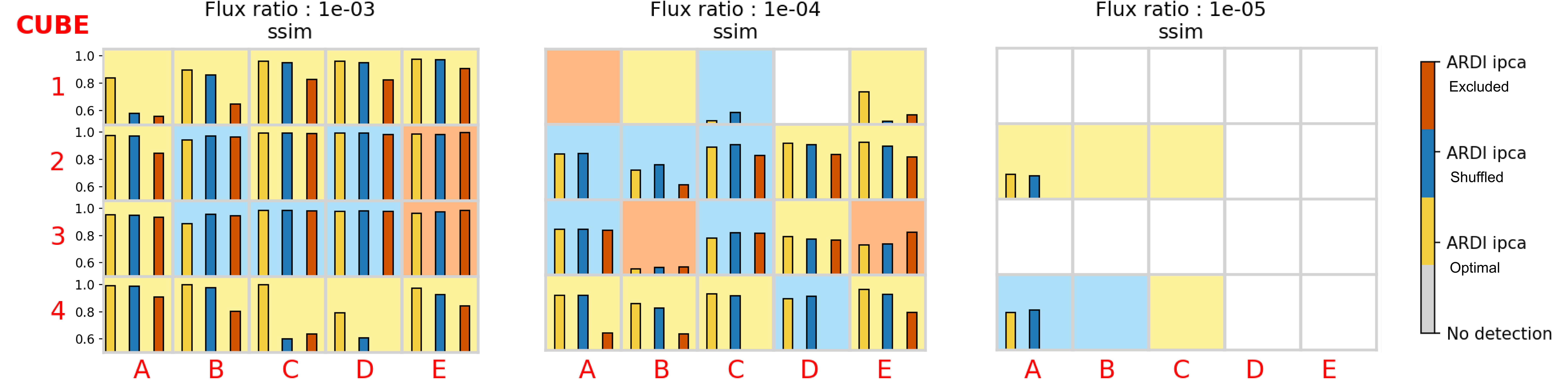}
        \caption{Results of our systematic tests comparing different qualities of references while utilizing ARDI for IPCA. The organization of this table is identical to Fig.~\ref{fig:results}, except that the value of the SSIM of the disk estimation compared to the ground truth is also represented for each algorithm within each cell.}
        \label{fig:results_mix}
\end{figure*}
A table representing the best algorithm for each dataset and for the three different strategies (ADI, RDI, and ARDI), is shown in Fig.~\ref{fig:results}. The quality of the extracted disk images is assessed using the SSIM metric. SSIM is chosen for its ability to cover multiple aspects of image similarity and its proven capacity to approximate structural differences akin to human visual perception \citep{SSIM}. In Appendix~\ref{ann:extratable}, we display the same table including bar plots in each cell to indicate the SSIM value obtained by each algorithm for each test case. However, given that assessing the quality of disk image estimation involves a multifactorial challenge, often requiring a balanced trade-off between minimizing geometric biases and reducing residual noise, we also consider four additional metrics -- namely, Pearson, Spearman, sum of absolute differences (SAD), and Euclidean distance -- in Appendix~\ref{fig:res_all}. The cases where all these metrics unanimously favor a specific algorithm are highly likely to stand out as clearly superior and are marked with a black star in the upper left corner of the table. The field of view considered for calculating the metrics affects the results. We calculated the metrics using a $1\arcsec$-radius aperture, excluding a 6-pixel radius inner circle located behind the coronagraphic mask. Complete results, including \remove{disk estimations}\new{the estimated disks ($\Bar{d}$}) for each algorithm and residual plots \remove{can be found in} in Appendix~\ref{ann:full_results} \new{are available on Zenodo\footnote{Appendix~\ref{ann:full_results} available at \url{https://zenodo.org/records/11442350}}}. The results for the second series of tests, aiming to compare three different selections of reference frames, are displayed in Fig.~\ref{fig:results_mix}, using a table with bar plots as in Appendix~\ref{ann:extratable}.

\begin{figure*}[!t]
    \centering
         \includegraphics[width=\textwidth]{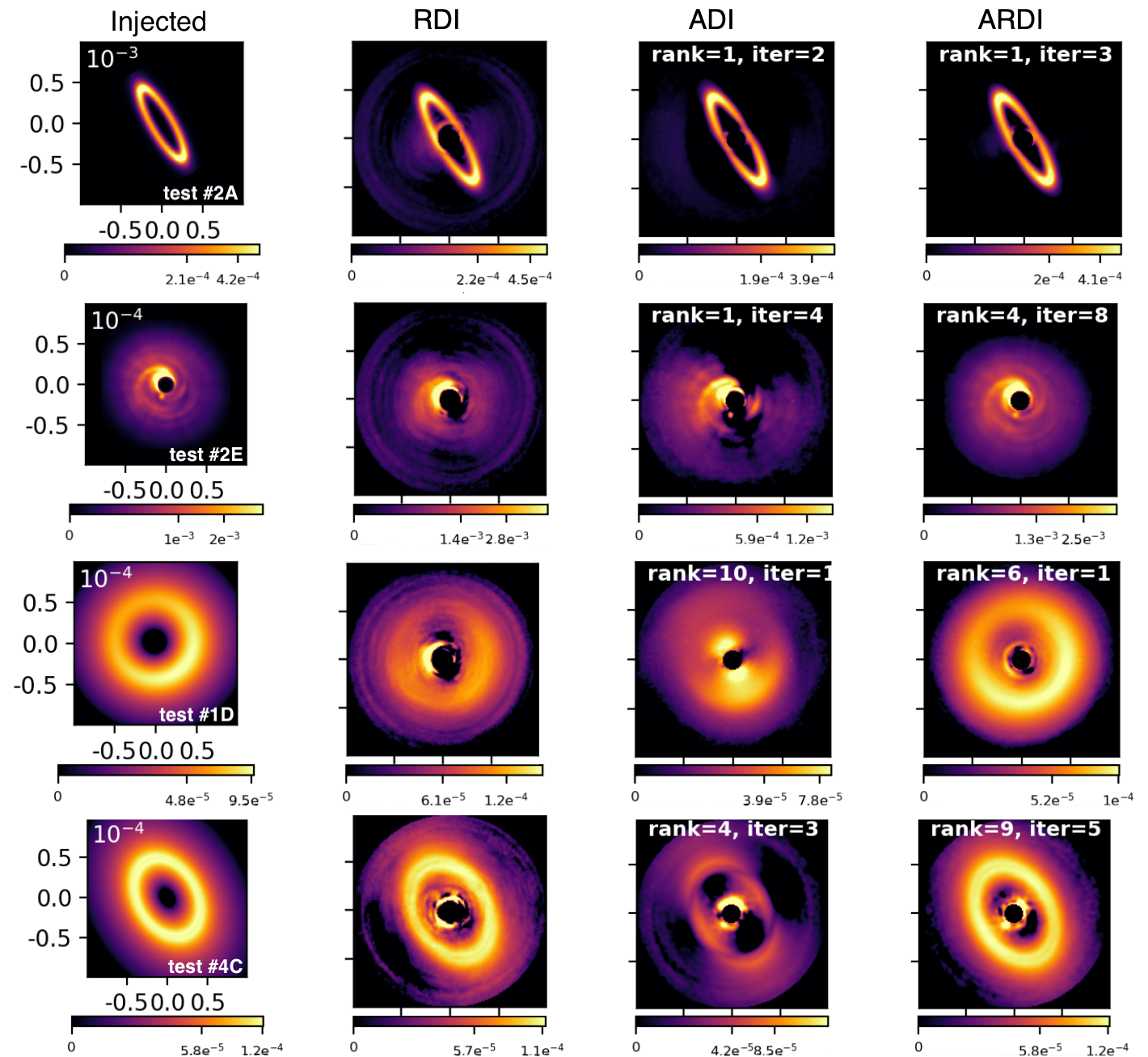}
        \caption{\nnew{Four examples of IPCA disk estimations obtained using RDI, ADI, and ARDI strategies, for a test dataset where ARDI performed significantly better than ADI and RDI. From top to bottom, the test dataset are cube \#2 with disk A injected at a contrast of $10^{-3}$, cube \#2 with disk E at a contrast of $10^{-4}$, cube \#1 with disk D at a contrast of $10^{-4}$, and cube \#4 with disk C at a contrast of $10^{-4}$. From left to right, the images display the injected disk and the results using RDI, ADI, and ARDI, utilizing the IPCA algorithm. These results were selected as they demonstrate ARDI's capability to compensate for the individual limitations of each strategy compared to using each method individually, in a variety of scenarios in terms of observing conditions, disk morphologies, and contrasts.}}
        \label{fig:selectionResults}

\end{figure*}

\subsection{Discussion of the test results}

The results of the first series of tests show that using the combined ARDI strategy, using optimal reference frames, is most often selected as the best strategy according to the SSIM (as well as other metrics). A significant visual improvement is observed for disk \#E, as shown in Figs.~{D.1E}, {D.2E}, {D.3E}, and {D.4E}, particularly at a medium contrast of $10^{-4}$. At this contrast, RDI struggles to capture the planet, while ADI has difficulty capturing the disk structure. \nnew{We show in Fig.~\ref{fig:selectionResults} the injected and retrieved disk images in four cases where ARDI performed significantly better than using either ADI or RDI individually.} The combination of both strategies proves to be the most relevant, as the weaknesses of each algorithm are effectively compensated for when used together. However, the degree of improvement provided by ARDI is not uniform across all datasets, including seven cases for which ADI or RDI alone perform better than ARDI according to SSIM. We propose to examine each of these specific cases:
\begin{itemize}
    \item ADI outperforms ARDI in dataset \#1 for disks A to C with a contrast of $10^{-4}$, as shown in Figs.~{D.1A}, {D.1B}, and {D.1C}. We observe that ADI achieves superior performance in scenarios where the dataset has a low Strehl ratio. Interestingly, this is also the same context where RDI shows lower performance. This observation suggests that ADI performs better when the information provided by the reference dataset is largely irrelevant. dataset \#1A (Fig.~D.1A) with a contrast of $10^{-3}$ is an exception, where ADI is selected as the best method based on the SSIM metric in Fig.~\ref{fig:results}. However, it is important to note that there is no consensus among the metrics. In fact, three of the five metrics (SAD, Euclidean and Pearson) favor ARDI, while one, Spearman correlation, indicates a preference for RDI. Upon visual inspection, it is obvious that ADI does not provide the most satisfactory result in this particular case.

    \item RDI seems to outperform ARDI in dataset \#4D and \#4E (Figs.~{D.4D}, {D.4E}) with a contrast of $10^{-3}$, as well as in dataset \#3A (Fig.~{D.3A}) with a contrast of $10^{-4}$. We note that RDI excells primarily in the case of face-on disks and favorable observing conditions. For instance, datasets \#4D and \#4E are almost perfectly recovered by RDI, suggesting that the angular diversity does not provide significantly more relevant information in these cases, as the information contained in the reference frames is sufficient on its own.
\end{itemize}

In terms of the restoration of the planet in disk \#E, we observe that at a contrast of $10^{-3}$, the planet is detected within the disk by all three strategies. However, ADI underperforms compared to the two other strategies, and the flux is not correctly retrieved as indicated by the error plots in Figs.~{D.1E}, {D.2E}, {D.3E}, and {D.4E}. However, at a contrast of $10^{-4}$, the planet is systematically and clearly detected with ADI, while with RDI no clear point-like feature stands out at the planet location. In fact, at a higher contrast, RDI struggles to capture the planet, while ADI has difficulty capturing the disk structure. ARDI captures the planet and the disk structure for cubes \#2, \#3, and \#4 but fails to recover both the disk and the planet for cube \#1. At lower contrast, neither ARDI nor ADI using IPCA seems to stand out for the task of planet detection only. Moreover, note that point-like artifacts can be observed in other synthetic disk estimates where no planets were injected, as in cube \#3 disk \#B at contrast $10^{-4}$ (Fig.~{D.1E}).

Working with a comprehensive battery of tests that require automation and standardization has the downside of potentially missing the best results that a particular algorithm could provide in specific cases. Indeed, while the behavior of these different strategies across a broader variety of scenarios remain relevant in most cases, a more specific optimization of the IPCA parameters might have significantly improved the quality of some of the disks, and potentially enabled more detections (e.g., by starting at a higher rank and reducing iteration per rank \remove{at higher rank} when processing the faintest disks of our test gallery).

%\subsection{Impact of reference frames selection}

In our second series of tests, we used the IPCA algorithm with three different selections of reference stars with different levels of quality according to PCC (see Fig.~\ref{fig:pearson_correlation}). The dataset \#4 (very stable) seems more sensitive to the selection of reference frames, while datasets showing more speckle variability (\#3, \#2, \#1) seem less sensitive. The mean PCC calculated for a nonoptimal selection is significantly lower for cube \#4 (mean PCC $\mu<0.6$) compared to other cubes (mean PCC $\mu\sim0.8-0.5$). 
As a result, the estimates produced when using these suboptimal references significantly under-performed, with SSIM hardly reaching 0.8 in most cases, while for other cubes, suboptimal references could reach similar SSIM as the optimal and semi-optimal selections (see Fig.~\ref{fig:results_mix}). Most interestingly, the three different selections achieved very similar results in terms of SSIM in many cases. In more than 40\% of cases, nonoptimal references achieved a higher SSIM than the optimal selection. The cases where the references contained 25\% of optimal references are very close to an optimal selection, even for a contrast of $10^{-5}$.
This suggests that selecting high-quality references, as assessed via PCC, has limited impact on the results in the final post-processed images using IPCA with ARDI. Instead, the outcomes appear to be more sensitive to the stability of the dataset.

\begin{figure*}[!t]
        \centering
       \includegraphics[width=\linewidth]{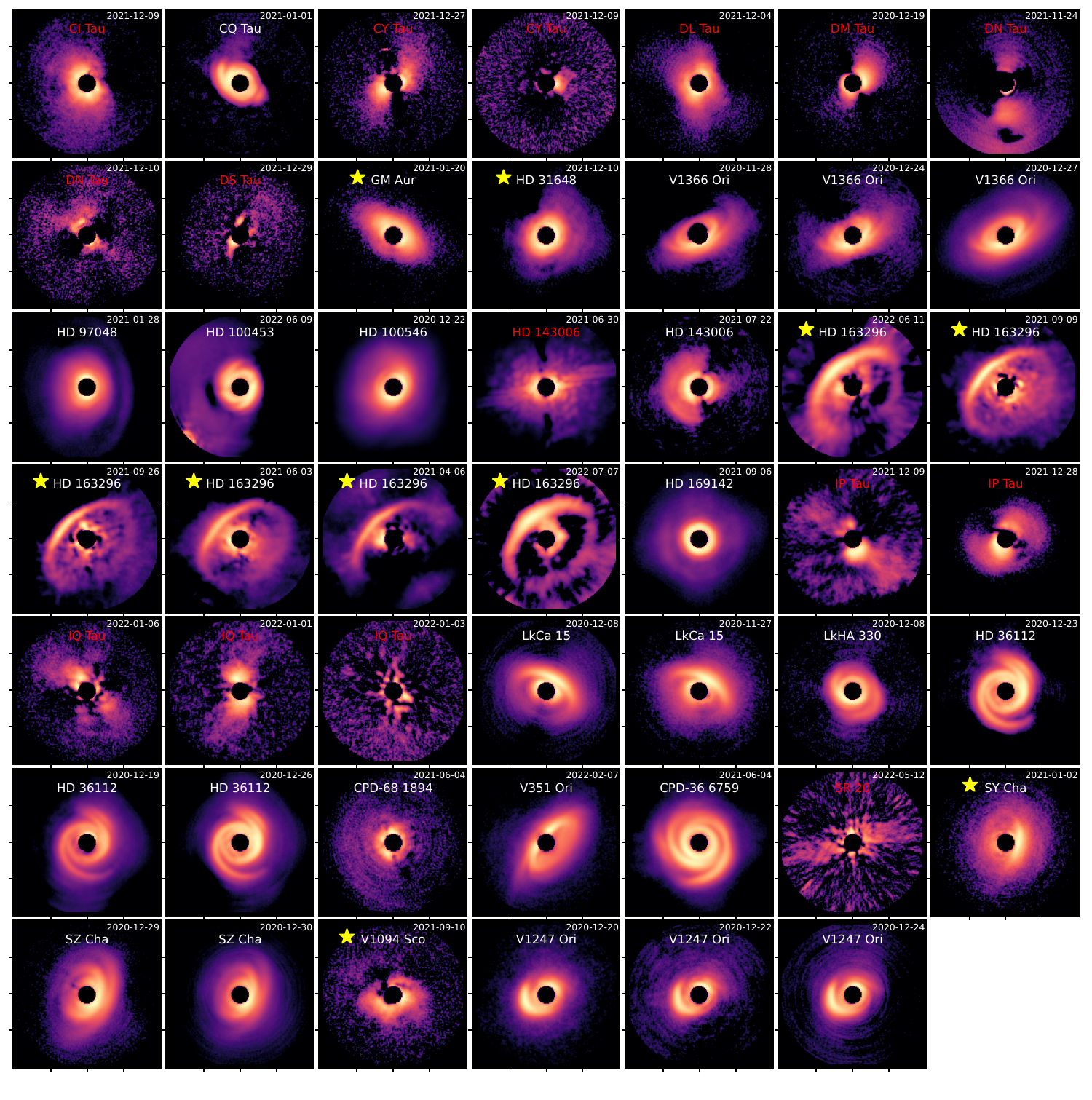}
        \caption{Gallery of protoplanetary disks datasets, originally described by \citet{Ren23a}, and re-processed here with IPCA-ARDI. The images have dimensions of $162\times 162$ pixels ($\sim 2\arcsec$) with logarithmic scale color bars. Images marked by a yellow star are those that were not presented in \citet{Ren23a} due to the low quality of the produced disk estimation, but that we still consider as a detection in our IPCA-ARDI results. Images marked with a red name are those that are not considered clear detection with IPCA-ARDI.}
        \label{fig:real_disks}
\end{figure*}

\section{Demonstration on protoplanetary disk datasets}
\label{sec:test_real}

To test our algorithm on real observations, we used the preprocessed datasets on protoplanetary disks from \citet{Ren23a}. The sample consists of 48 datasets including 29 \new{unique} different young stars hosting protoplanetary disks with known substructures previously observed in scattered light. Among them, four were reported to have exoplanet candidates (or claims thereof) from high-contrast imaging: HD 100546 \citep{Quanz2013,Quanz15, Currie14}, HD 169142 \citep{Biller14, Reggiani14, Gratton19,Hammond23}, LkCa 15 \citep{Kraus12, Sallum15}, and MWC 758 \citep{Reggiani18, Wagner19, Wagner23}. Additionally,  localized deviations from Keplerian rotation possibly associated with the presence of planets were reported in HD 163296 \citep{163296KINK, Pinte20, Pinte18} and HD 97048 \citep{KINKHD97048}.  All observations were made in the broad-band $Ks$ filter using the IRDIS camera of VLT/SPHERE. These disks were previously presented by \citet{Ren23a} in both polarized light using the \texttt{IRDAP} \citep{irdap} pipeline for polarimetric data processing, and in total intensity using RDI through the DI-sNMF algorithm. The references were acquired via the ``star-hopping'' mode \citep{Wahhaj21}, which means that the telescope is regularly pointed at a nearby reference star during data acquisition, efficiently alternating science and reference observations with minimal overhead. This observing strategy ensures maximal similarity between the respective PSFs.

We present our reprocessing of the \citet{Ren23a} data with IPCA using the ARDI strategy in Fig.~\ref{fig:real_disks}. For most datasets, a reference frame ratio ($r/n$) of one was employed, utilizing the dedicated reference star observed in star-hopping mode. The default parametrization method involves ten iterations per rank, starting from rank one to rank ten. However, for some datasets, the parametrization was individually optimized. In cases where the signal was faint, the starting rank was set to a higher value, and the number of iterations per rank was smaller. In situations where the angular diversity was too small, we increased the reference frame ratio. When reference stars were unsatisfactory (e.g., for HD 163296), we employed a combination of the most correlated references available from the references acquired in star-hopping mode from this sample of protoplanetary disk observations. In these cases, we also utilized a higher reference frame ratio ($r/n$) and used a higher starting rank. We manually selected the best image from the estimates generated for these parameters. IPCA parameters and the selected frame for each dataset can be found in Appendix~\ref{sec:paramIPCA}. \new{Table~\ref{tab:params} also provides the mean and standard deviation of the PCC between the science data and the references, following the same procedure detailed in Sect.~\ref{sec:testsim}.} 

\remove{In their study,} \new{Among the 48 datasets}, \citet{Ren23a} showcased high-quality results for \new{23 datasets\footnote{Only 18 datasets processed with DI-sNMF are presented in \citet{Ren23a}, but among the non-presented ones, five more can be considered as clear detections (private communication).} including 15 unique systems} \remove{among the 48 datasets }using DI-sNMF, selectively discarding instances of non-detections and very low quality images. In the context of this article, we present all results processed with IPCA-ARDI, explicitly including the non-high quality images. Among these images, 33 \new{datasets including 19 unique} protoplanetary disks are visually recovered, exhibiting a range of image qualities from excellent (e.g., V351 Ori, HD 36112, or V1366 Ori) to bare detection (e.g., GM Aur or HD 163296). This variability is influenced by factors such as the quality of the references, the brightness of the disk, and the degree of angular diversity. Comparing the results, it is noteworthy that for the 18 datasets analyzed with both DI-sNMF (RDI) and IPCA (ARDI) detections, similar disk structures are observed.

\subsection{Inspecting protoplanet claims in individual systems}

\begin{figure}[!t]
    \centering
    \includegraphics[width=0.92\linewidth]{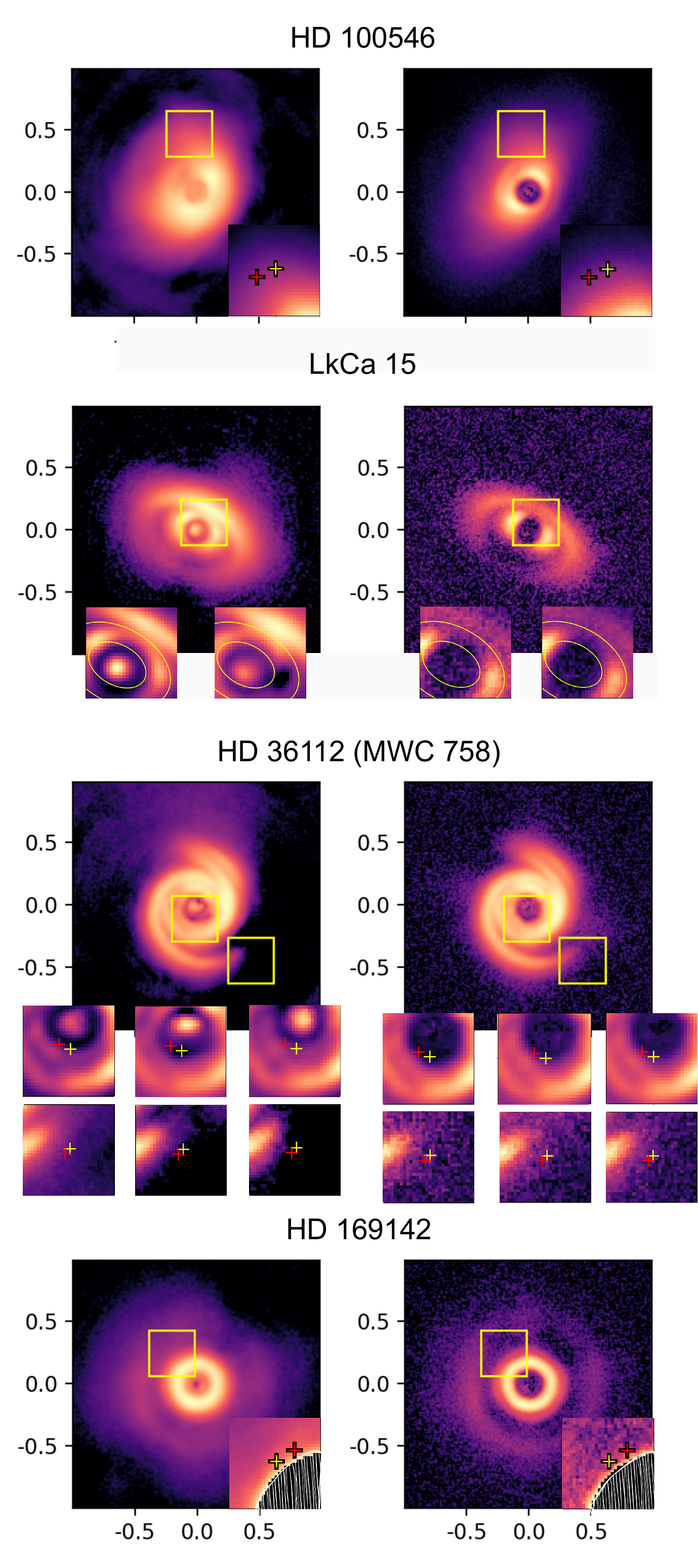}
    \caption{ARDI images processed with IPCA (left) and polarized images processed with IRDAP (right) for all protoplanetary disks containing a planet claimed via direct imaging (in logarithmic scales). An approximate position of the candidate is marked with a yellow square on the full image, and a zoom on the candidate's position is shown at the bottom ($30\times 30$ pixels, linear scales). In each sub-image, the yellow cross marks the candidate's approximate position in the discovery paper, and the red cross the expected position on the date of acquisition of the SPHERE images presented in this work, considering a circular orbit. For LkCa 15, given the multiple claimed point-like features at varying positions, we only show two yellow orbits at 10 and 30\,au defining the region where the candidate(s) were proposed.}
    \label{fig:planetdisk}
\end{figure}

\begin{figure}[!t]
    \centering
    \includegraphics[width=\linewidth]{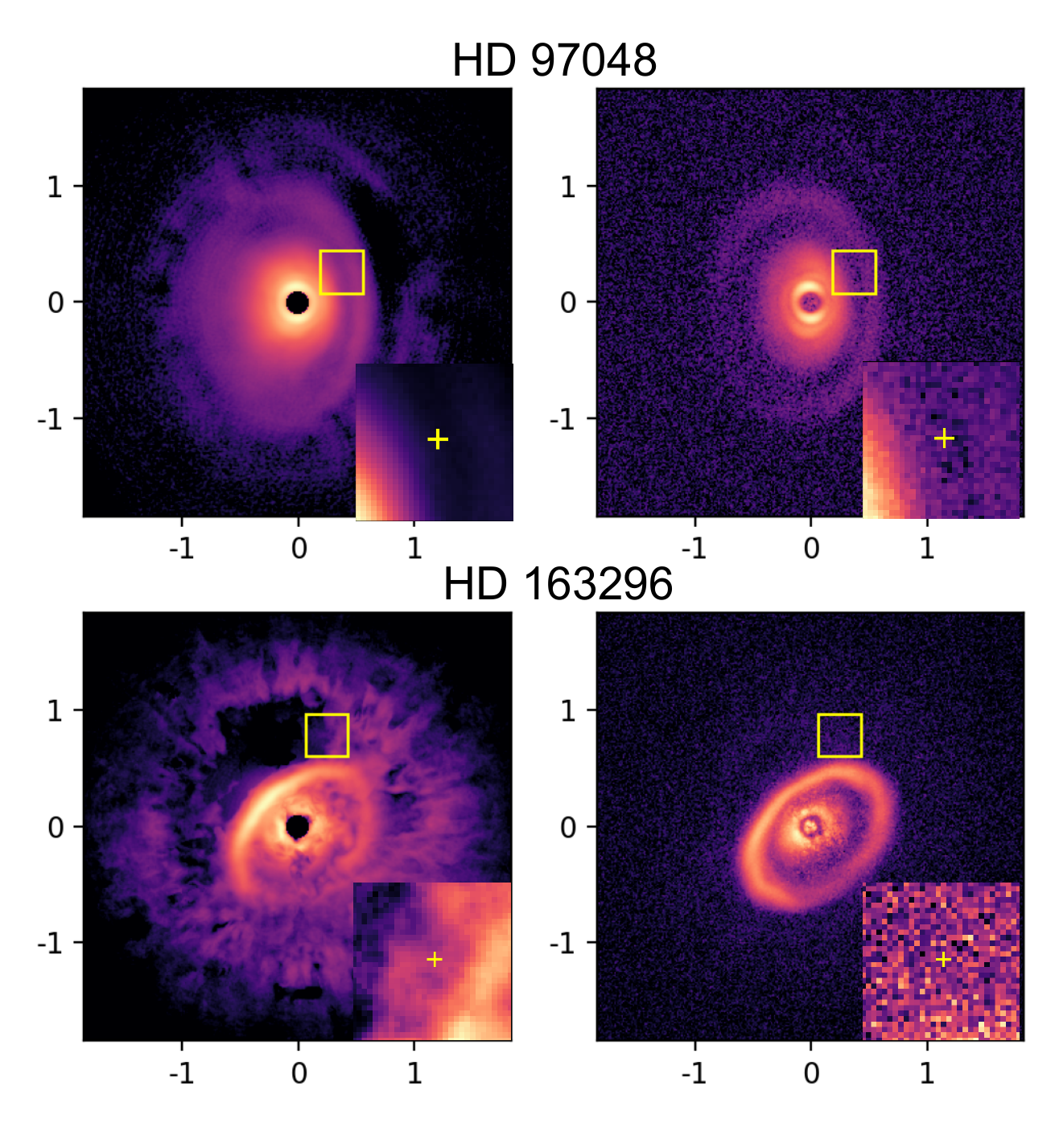}
    \caption{Same as Fig.~\ref{fig:planetdisk}, but for protoplanetary disks containing a planet claim deduced via the observation of local deviations in Keplerian velocity. }
    \label{fig:kinks}
\end{figure}

Among the systems presented above, four have unconfirmed protoplanet claims from direct imaging observations \new{(including two such claims for MWC 758)}, and two exhibit kinematic signatures that could indicate the presence of planets \new{(including two candidates for HD 163296)}. The different images showing the position of the candidate(s) within their protoplanetary disk are provided in Figs.~\ref{fig:planetdisk} and \ref{fig:kinks}. We propose to review the different candidates in the following paragraphs.

\paragraph{HD 100546.} A point-like feature, interpreted as a potential gas giant planet at 52\,au from its star, was observed in 2013 with the NACO instrument installed at the VLT in the $L'$ and $M'$ bands \citep{Quanz2013, Currie14, Currie15, Quanz15, Currie17}. These observations were processed using the ADI strategy with the PCA and LOCI algorithms. The potential candidate is located within the broad-disk signal, which was not recovered in the processed images shown in these publications. Indeed, the disk was filtered out due to the now well-known self-subtraction and over-subtraction effects \citep{Sandrine23, Sandrine, Milli}, which worsen when using a large number of principal components. \new{Considering the expected Keplerian motion on a circular orbit between 2013 and 2020, the candidate would have moved by approximately 8\degr\ in position angle, which corresponds to a 6.5-pixel shift in the image (see Fig.~\ref{fig:planetdisk}). No companion is recovered in our images at those positions.} In the ARDI images, as well as the DI-sNMF results, an extended spiral-like structure stands out rather than a point-like feature. \new{Due to the presence of strong disk signal dominating in the area where the candidate is located, we cannot provide robust detection limits in the $Ks$ filter for this candidate.}

\paragraph{LkCa 15.} Potential candidates were found using observations in $L'$ and $K$ bands taken with the near-infrared imager NIRC2 of the Keck II telescope in 2009 and 2010 \citep{Kraus12}, followed by several observations between 2014 and 2015 using the Large Binocular Telescope in $L'$, $Ks$, and H$\alpha$ filters \citep{Sallum15}. The detections were interpreted as potential newly formed gas giant planets surrounded by dusty material. However, it was mentioned that the observed signal could represent a more complex structure rather than a single point source \citep{Kraus12}. In fact, candidates buried within the disk pose a significant challenge for detection. Additionally, limitations arise due to the high variability of the speckle field in that specific image region and due to the small working angle. Recent observations of the disk support the hypothesis of a filtered disk signal, as there are no discernible point-like features (PLF) in the disk \citep{Currie19}. The images provided by both DI-sNMF and IPCA-ARDI corroborate this statement, since no PLF is observable. The second row in Fig.~\ref{fig:planetdisk} presents the approximate positions where the diverse candidates should be located, displaying two bound orbits at 10 and 30\,au from the star. \new{Inferring detection limits for these candidates would be highly unreliable, as they are located within an area of strong residual signals likely tracing a combination of inner disk and stellar residuals.} % Additionally, there is a significant disparity in signal level within the annulus where the planets are located between our two datasets, acquired within only a 2-week interval.

\paragraph{HD 36112 (MWC 758).} A first bright candidate was reported by \citet{Reggiani18} based on data acquired in 2015 and 2016 using Keck/NIRC2 in $L'$ band. The candidate, located at $0\farcs11$ from its star, was captured from two ADI datasets processed using PCA. The possibility that a bright artifact appeared twice at the same location was considered, with an odds ratio of approximately $\sim$1/1000. In the vicinity of the reported candidate, the new images presented in this article and in \citet{Ren23a} reveal a complex disk structure. However, it remains unclear whether these structures are related to the PLF reported by \citet{Reggiani18}, which was notably brighter than the rest of the disk. An in-depth analysis is deferred to a future publication (Christiaens et al., in prep.). Additionally, a second, very red protoplanet candidate was claimed by \citet{Wagner19, Wagner23} near the tail of the northern spiral using observations from 2016, 2017, 2018, and 2019 with LMIRCam mounted on the Large Binocular Telescope Interferometer. Further analysis of these two candidates, carried out by \citet{Boccaletti21}, did not re-detect the candidates proposed by either \citet{Reggiani18} or \citet{Wagner19}, and did not find them consistent with being spiral-driving planets assuming linear density wave models. \new{For these two candidates, the expected motion between the first observation and the SPHERE Ks-band observations (shown in Fig.~\ref{fig:planetdisk}), according to Keplerian motion on a circular orbit, would be respectively a $3.9$-pixel shift and a $2.2$-pixel shift.} We do not re-detect them in our IPCA-ARDI images, nor did \citet{Ren23a} using DI-sNMF. We performed planet injection \new{at the separation of the \citet{Wagner19} candidate companion}, and generated a contrast curve using PCA with ADI-only in Appendix~\ref{ann:36112}. \new{Our analysis reveals a detection threshold of approximately $2.6 \times 10^{-5}$, which is consistent with the contrast limit established by \citet{Grady13} at this separation. According to the ATMO2020 model \citep{Phillips20} and assuming no extinction, our contrast threshold translates into a 1 Jupiter mass ($M_{\text{Jup}}$) sensitivity. This mass sensitivity does not matches with the predicted mass of about 2-3$M_{\text{Jup}}$ inferred by \citet{Wagner19}. This tension could be explained by the usage of different planet evolution models, or possibly by extinction or by the presence of a circumplanetary disk (more details in Appendix~\ref{ann:36112})}. \new{Converting the contrast of $1 \times 10^{-5}$ for the candidate in the $L'$ band as presented in \citet{Wagner19} into a mass using the ATMO2020 model for a consistent comparison, we find a mass of $0.5 M_{\text{Jup}}$, suggesting that our detection limit would not allow us to detect the candidate.}
    
\paragraph{HD 169142.} \citet{Reggiani14} and \citet{Biller14} independently observed a PLF in $L'$ band using independent ADI datasets from VLT/NACO %and Gemini/GPI, 
processed with PCA. %\citet{Reggiani14} suggested the presence of a low-mass companion, whether a brown dwarf or a forming planet. Instead, \citet{Biller14} suggested that it must be a disk feature, given the absence of re-detection in the H or Ks band despite the candidate's relative brightness. Indeed, 
Better images of the disk obtained with SPHERE suggested that this candidate at $0\farcs11$ separation may rather trace filtered signal from the inner ring of the disk \citep[e.g.,][]{Ligi19}. Later, another candidate was proposed using observations in polarized light from 2015, along with two ASDI observations in 2017 and 2019 using the IRDIS and multiple IFS observations between 2015 and 2019 \citep{Gratton19,Hammond23}. %A point-like feature was observed in the cavity in both polarized and ASDI images. 
However, detecting the candidate with ASDI required suppressing most of the disk signal. The observations were conducted in $J$ and $YJH$ bands and the position of the candidate between 2015 and 2019 appeared to follow a Keplerian motion. The signal was interpreted as a gap-clearing Jovian-mass protoplanet surrounded by a circumplanetary disk or envelope. %a potential protoplanet embedded with a circumplanetary disk or envelope surrounding a gap-clearing Jovian-mass protoplanet. 
Studies by \citet{Garg21} and \citet{Law23} using ALMA observations revealed several distinct chemical signatures that demonstrate a compelling link to ongoing giant planet formation. Furthermore, compact CO J=2–1 and CO J=3–2 emission counterparts were observed that coincide with the location of the candidate planet. %\new{From the initial observation in 2015 to the observation considered in this work (2021), the expected Keplerian motion would result in approximately a 6.7-pixel shift.}
Yet, no PLF appears in polarized light nor in total intensity for both the DI-sNMF and IPCA-ARDI of the $K$-band images presented here\new{, at its predicted location at the epoch of the Ks observations considered in this work (2021) assuming Keplerian motion}. In Appendix~\ref{ann:169142}, we attempt to infer the detectability of the candidate by injecting a fake planet at the same separation, using the estimated contrast in \citet{Hammond23}. In this test, we processed the dataset with IPCA-ARDI as well as plain PCA-ADI. In addition, we provide the contrast curve obtained via PCA-ADI in Fig.~\ref{fig:cc169142}. \remove{The results suggest that the companion is not detectable in our images using PCA-ADI and IPCA-ARDI techniques, considering the flux and separation reported in Hammond23}. \new{Our $5\sigma$ detection limit of $1 \times 10^{-4}$ at the expected location of the companion translates into $2 M_{\text{Jup}}$ according to the ATMO2020 model, assuming no extinction. In comparison, \citet{Hammond23} report a $YJH$ contrast of $\sim 1.5 \times 10^{-5}$. Nevertheless, the detection of polarized intensity colocated with the protoplanet suggests the measured $YJH$ light is dominated by scattered light, making the mass conversion using atmospheric models most probably irrelevant. Assuming similar expected contrast in $Ks$ band as in $YJH$, our senitivity limits do not allow us to confirm or refute the candidate.}

\paragraph{HD 97048.} A kinematic signature was detected in the gap of the disk around HD 97048 using high-spectral and spatial-resolution ALMA observations \citep{KINKHD97048}. This signature could be explained by the presence of a superjovian planet in the disk, possibly at 130\,au from the star. \new{At this separation, the motion of the candidate between the date of the first detection and our dataset would have been less than a 1-pixel shift.}  Unfortunately, we were unable to directly observe any point-like feature at this location in total intensity. We injected fake planets at the same separation as the candidate to assess the contrast sensitivity of our reduction at its separation, and found that a candidate below a contrast of $1 \times 10^{-5}$ in $Ks$-band would not have been detected. \new{Based on our contrast limit, the ATMO2020 models imply an upper mass limit of $0.5 M_{J}$ (see Appendix~\ref{ann:97048}). The discrepancy with the kinematically inferred mass may suggest the presence of extinction or the inadequacy of the ATMO2020 models for embedded protoplanets.}
        
\paragraph{HD 163296.} Using CO line observations, \citet{Pinte18} initially reported a local deviation from Keplerian velocity, suggesting the presence of a massive body at approximately 260\,au from the star. %A second deviation around 86 AU was also mentioned in \citet{Pinte20}. However, it was not resolved in velocity (i.e., not detected in at least three independent channels), meaning that no link to a candidate planet could be established. 
Subsequently, \citet{163296KINK} reobserved the first candidate proposed by \citet{Pinte18}, along with a second localized deviation from Keplerian rotation, potentially associated with the presence of a giant planet at 94\,au. The first candidate proposed by \citet{Pinte18} and \citet{163296KINK} at 260\,au does not fit within the field of view of the images presented in Figs.~\ref{fig:kinks} and \ref{fig:real_disks}. \new{At these separations, the motion of the candidates between the date of the first detection and the new dataset is negligible.} We show another post-processed image for the same dataset, with an uncropped field-of-view in Appendix~\ref{ann:163296} as well as contrast curve with PCA-ADI. No PLF is observed at the locations of the two proposed sources in these direct imaging observations with both IPCA-ARDI and PCA-ADI, despite the estimated sensitivity reaching down to $3\times 10^{-4}$ in contrast \new{(i.e., mass limit of $1M_J$) for the candidate at 260\,au, and $6\times 10^{-6}$ in contrast (i.e., mass limit of $4M_J$) at 94\,au (see details in Appendix~\ref{ann:163296})}.

\paragraph{}The direct observation of a protoplanet candidate within a protoplanetary disk in direct imaging can necessitate \remove{significant trimming of the disk signal}\new{the use of aggressive post-processing techniques that lead to significant disk flux loss and alterations to the apparent morphology of the disk in order to reveal faint point-like sources.} \remove{through post-processing to reveal faint peaks}. Many of the directly imaged protoplanet claims presented in this sample rely on poor-quality images, exhibiting signs of self- and over-subtraction. It remains unclear whether a PLF results from filtered disk signal, from artifacts inefficiently corrected by post-processing, or from an actual planet, especially for candidates buried within the disk signal. This ambiguity makes it challenging to both confirm and refute these candidates. Indeed, the absence of the proposed candidates cannot definitively discard the claims, as the wavelengths might be inappropriate, the contrast could be too high, or the candidate might be buried within the disk. Being able to capture high-quality images of the disk using techniques such as ARDI with IPCA is still relevant to prevent false candidates due to filtered disk signals.

% \begin{figure}[t]
%     \centering
%     \begin{subfigure}[b]{0.45\textwidth}
%         \centering
%         \includegraphics[width=\textwidth]{Images/planetdisk.drawio.png}
%         \caption{}
%         \label{fig:planetdisk}
%     \end{subfigure}
%     \hfill
%         \begin{subfigure}[b]{0.45\textwidth}
%         \centering
%         \includegraphics[width=\textwidth]{Images/kinks.png}
%         \caption{}
%         \label{fig:kinks}
%     \end{subfigure}
%             \caption{Kinks and planets. Need a caption}
% \end{figure}

% The problem of parameter choose is common to the PCA algorithm widely used for direct imaging of exoplanet. Methods that analyses the PCA results though statistical tools, such as NFM or "" can assist users for this task, despite their heavly computational cost. As for exemple, the RSM algorithm does utilize statistical tools to leverage the best planet estimation from speckle field estimate via PCA. Through the exoplanet high-contrast daa chanllenge condcuted in \citep{}, RSM have been show efficient compare to other method. It is possible to envisage adapting such tools for the usage of disk imaging via IPCA.

% Regarding parameter optimization, one interesting observation is that the convergence of significant signals, as shown in Figure \ref{fig:ARDIconv} and Figure \ref{fig:ADIconv}, suggests that the information coming from the reference stars enables more disk regions to converge properly. In ADI alone, we observed that noise regions or regions where invariance to rotation is significant do not converge properly through the iterations.

\section{Conclusion}

We explored the possibility of combining the RDI and ADI observing strategies by adapting the IPCA algorithm originally designed for ADI datasets. Our main goal was to assess the extent to which combining information from ADI and RDI could enhance disk \remove{estimation}\new{recovery} compared to using each strategy independently, while employing IPCA-based methods. Our analysis encompassed systematic testing on a diverse set of 60 synthetic datasets, covering a range of observing conditions and disk morphologies. We tested our methods with both optimally chosen reference stars and with a random mix of reference stars to assess the impact of reference star quality on the estimated disk.

Our results revealed that ARDI consistently improves the recovery of extended signals compared to using these techniques individually. This improvement holds true across various scenarios, encompassing different observing strategies and disk morphologies. A significant visual improvement from ARDI occurs when a planet is hosted within a disk. At lower contrast levels and for the more variable speckle pattern, RDI struggled to capture the planet, while ADI had difficulty capturing the disk structure\new{, particularly due to the rotation-invariant flux, which is} an inherent ambiguity of the observing strategy\remove{: the rotation-invariant flux}. The combination of these strategies emerges as a reliable approach to mitigate the limitations of individual observing strategies, addressing challenges that would otherwise be difficult to overcome.

We also explored the problem of parameters' optimization for IPCA algorithms. While a good choice of parameters is important, we found that multiple sets of parameters \remove{should} lead to similar results. Nevertheless, we identified different scenarios and matched them with specific parametrizations to provide optimal results in each case. For the brightest disks with a stable speckle field, it is better to start the iterative process at rank one and have multiple iterations per rank before increasing the rank (e.g., CPD-68 1894 or HD 36112 data presented in Sect.~\ref{sec:test_real}). This ensures that most of the self-subtraction effects have been mitigated before loosening the model of the speckle field by increasing the rank. Conversely, when the disk is faint relative to the speckle, it is necessary to start at a high rank and limit the number of iterations per rank to avoid the bright artifacts stemming from non-captured speckle field components from being propagated in the subsequent iterations (e.g., V1094 Sco or GM Aur datasets). Regarding the convergence of the iterative process, we acknowledge the difficulty in building a reliable stopping criterion, which is a common problem in computer science \citep{Nocedal2002}. It is particularly challenging in our case of application, as determining where to stop the iterative process involves choosing a good compromise between noise subtraction (non-propagation) as well as the filtering and recovery of extended signals (convergence), and also because different parts of the disk converge at different rates. Nevertheless, observing the convergence curves remains an interesting indicator of the optimal iteration to be reached.

%Ensuring a few degrees of angular diversity during data acquisition ($>10\deg$), would facilitate the exploitation of data through ARDI, thereby mitigating potential challenges arising from reference stars. Hopefully, datasets acquired in star hopping mode inherently present a good amount of angular diversity rotation; which make of ARDI a suitable alternative to many archival datasets.

We applied our IPCA algorithm with ARDI to reprocess a sample of 48 datasets that includes 29 different young stars surrounded by protoplanetary disks exhibiting known substructures, which were initially introduced by \citet{Ren23a} in both polarized light and total intensity, using RDI through the DI-sNMF algorithm. All disks from the 29 young protoplanetary disk gallery published by \citet{Ren23a} that were successfully recovered by DI-sNMF (RDI) were also successfully retrieved with IPCA-ARDI. Under optimal conditions with ideal reference stars, IPCA demonstrates capabilities comparable to those of DI-sNMF in preventing over-subtraction. Furthermore, the usage of ARDI enables the detection of fainter disks such as V1094~Sco.

While a few planet candidates were reported via direct imaging among the sampled young stars, we did not reobserve any of these candidates. \new{This is at least partly due to limitations in the achieved sensitivity in the considered filter ($Ks$), which renders the detection of these companions challenging or unfeasible.} While confirmed detections of protoplanets within disks remain rare, with only a select number being reported in the literature \citep[e.g.,][]{PDS70info, Haffert19}, studying young systems in their formative stages embedding protoplanetary disks housing nascent planets is crucial for understanding the intricate process of planetary system formation. The application of IPCA-ARDI holds promise in providing valuable insights by preserving both the structures of the disk and clear images of planets.

%% ---------------------------- REFS and ABSTARCT ------------------------------------------- %%

\begin{acknowledgements}
This project has received funding from the European Research Council (ERC) under the European Union's Horizon 2020 research and innovation program (grant agreement No. 819155), and from the Belgian Fonds de la Recherche Scientifique -- FNRS. This work has made use of the SPHERE Data Centre, jointly operated by OSUG/IPAG (Grenoble), PYTHEAS/LAM/CESAM (Marseille), CA/Lagrange  (Nice), Observatoire de Paris/LESIA(Paris), and Observatoire de Lyon \citep{Galicher, Delorme17}. We are very grateful to Bin Ren and Myriam Benitsy for sharing their data used in this work, and to Philippe Delorme, and Mariam Sabalbal for their help in selecting, providing, and pre-processing the SPHERE data cubes used for the tests. 
SS acknowledges funding from the European Research Council (ERC) under the European Union's Horizon 2020 research and innovation program (COBREX; grant agreement n° 885593).
\end{acknowledgements}

\bibliographystyle{aa} % style aa.bst %BR: I hacked the citation files and this would allow direct clicking on the references
\bibliography{references}

\begin{appendix}

\section{Local convergence of IPCA-ARDI}
\label{ann:patches_gradient}

\begin{figure}[!h]
    \centering
    \includegraphics[width=0.9\linewidth]{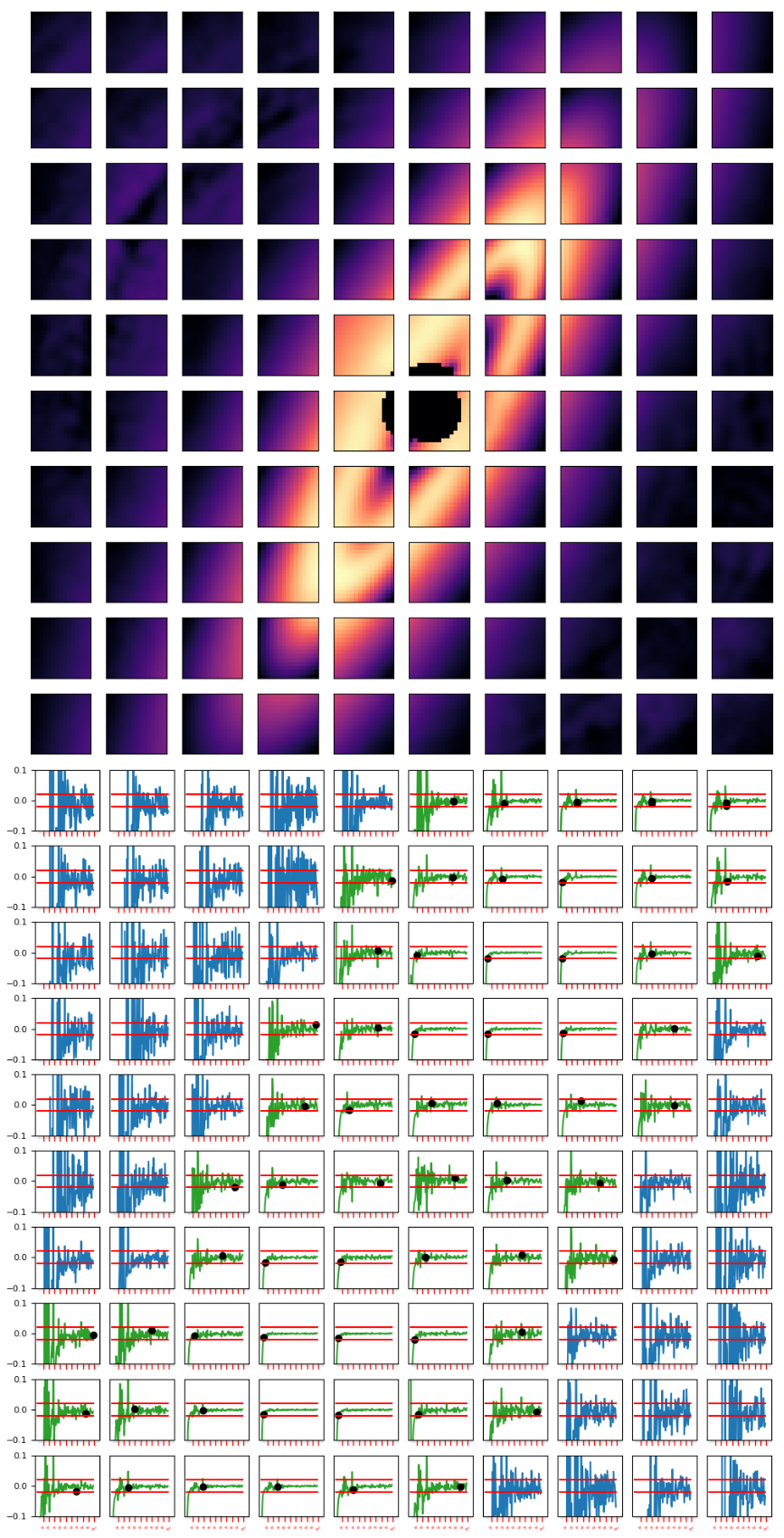}
    \caption{Evolution of the IPCA with ARDI estimate through the iterations. The dataset used is the same as in Fig.~\ref{fig:ARDIconv}. The first grid of plots (top) displays the final estimation segmented into patches of size $15 \times 15$ pixels. The second grid of plots (bottom) corresponds to the gradient computed within these patches according to Eq.~\ref{equ:gtol}. The red lines indicate the $g_{\rm tol}$ threshold at $\pm 10^{-2}$. If the threshold is reached, the curves are displayed in green, and a black dot indicates at which iteration the signal has converged. Otherwise, the curves are displayed in blue.}
    \label{fig:patches_gradient}
\end{figure}

In this appendix, we study the evolution of the integrated flux into different patches of the image through the IPCA iterative process when using ARDI. Figure~\ref{fig:patches_gradient} shows that the brightest part of the disk converges faster than the rest of the image. We can also observe that the parts of the image where no signal is present do not seem to converge. It is important to note that the observed non-convergence on the outer parts of the image where there is no disk signal relates to the expression of the gradient. Indeed, from Eq.~\ref{equ:gtol} it is clear that for an estimate $d_i$ that tends toward zero, the quantity $\frac{d_i-d_{i+1}}{d_i}$ becomes more sensitive to minor variations.
\vspace{20pt}

\section{Comparison of the SSIM score obtained by each algorithm}
\label{ann:extratable}

\begin{figure*}[!t]
    \centering
         \includegraphics[width=0.9\textwidth]{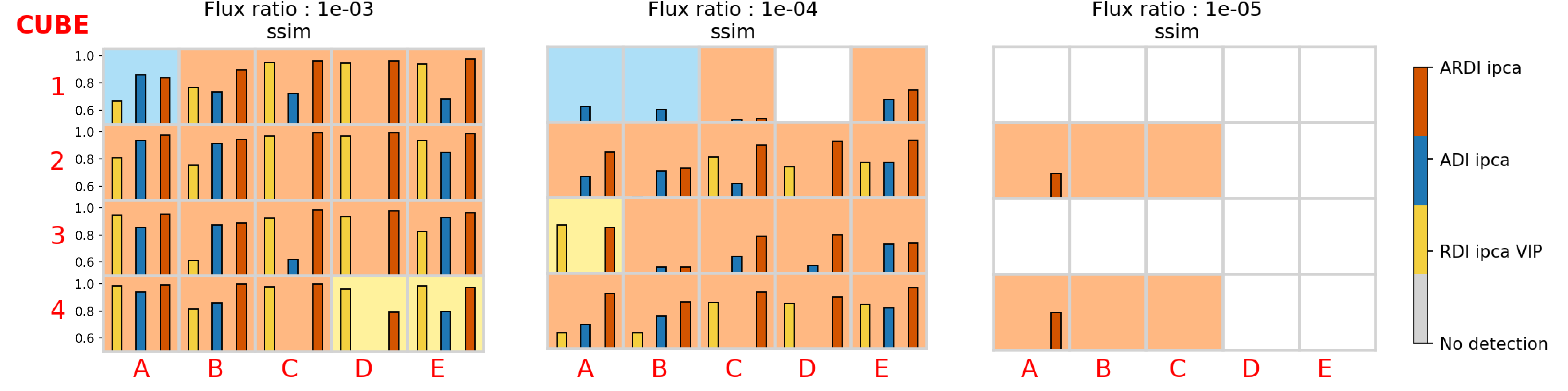}
        \caption{Results of our systematic tests comparing different strategies, namely RDI, ADI and ARDI \remove{while}\new{with the} IPCA algorithm. The organization of this table is identical to Fig.~\ref{fig:results}, except that the value of the SSIM of the estimation obtained by each algorithm compared to the ground truth is also represented within each cell by a colored histogram with RDI in yellow, ADI in blue, and ARDI in red.}
        \label{fig:resultsbar}
\end{figure*}

This appendix presents the results of the performance comparison for the three considered strategies (RDI, ADI, ARDI) using the IPCA algorithm, for each test dataset, using the structural similarity index measure \citep[SSIM,][]{SSIM}. For each of the three $4\times 5$-sized tables in Fig.~\ref{fig:resultsbar}, each cell represents a different test dataset (see Sect.~\ref{sec:test} and \citet{Sandrine23} for more details about the test datasets). In each cell, three bar plots represent the SSIM obtained between the ground truth and the final disk image obtained with each strategy using the IPCA algorithm. White cells indicate that no strategy detected the disk.

\section{Comparison of results for different metrics }
\label{sec:winners}

\begin{figure*}[!t]
    \centering
    \setlength{\belowcaptionskip}{-5cm}
    \includegraphics[width=0.85\linewidth]{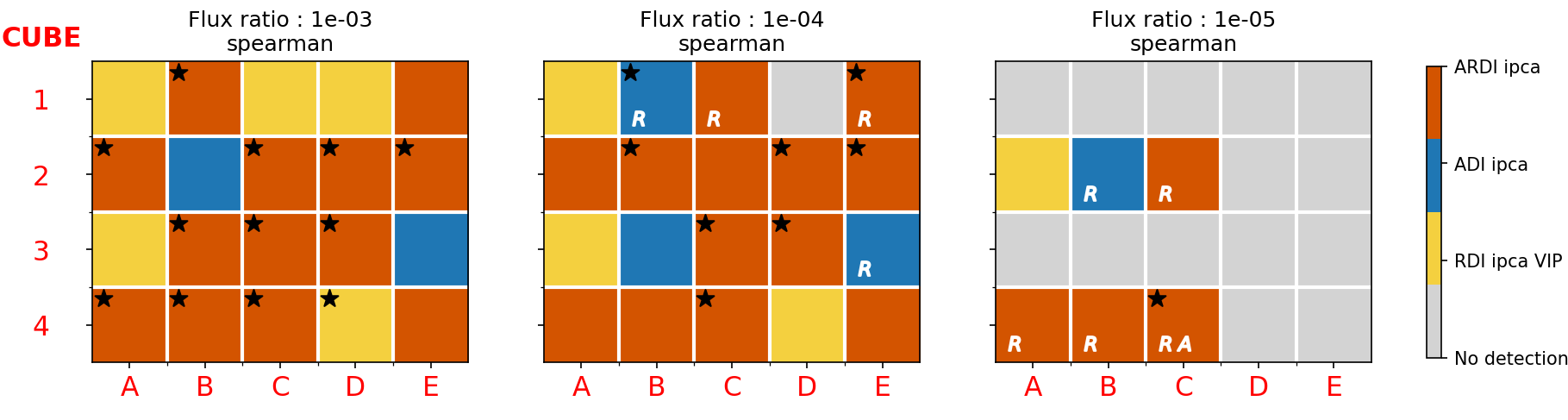}
    \includegraphics[width=0.85\linewidth]{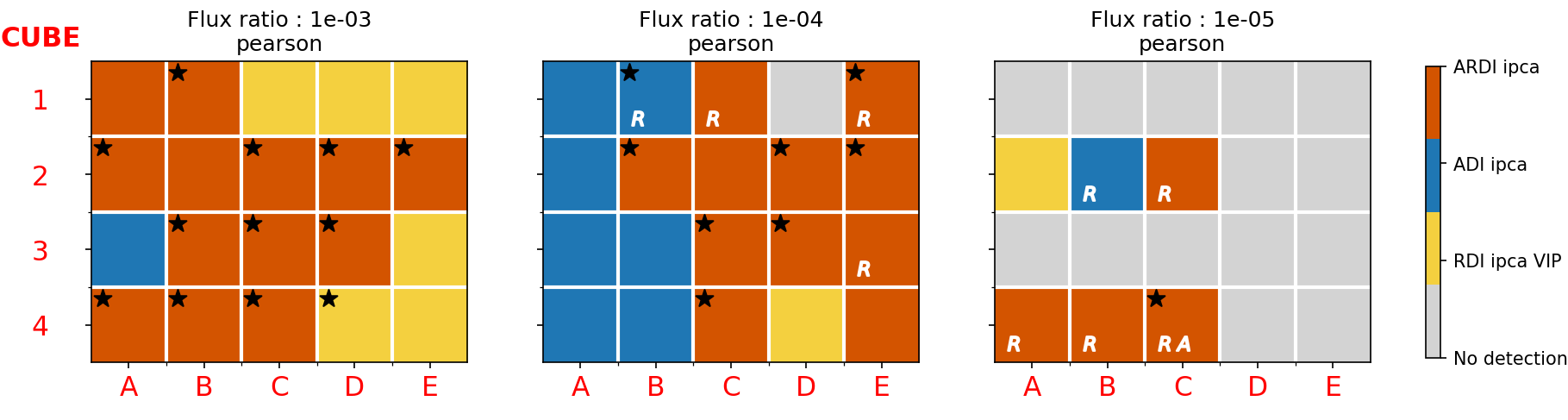}
    \includegraphics[width=0.85\linewidth]{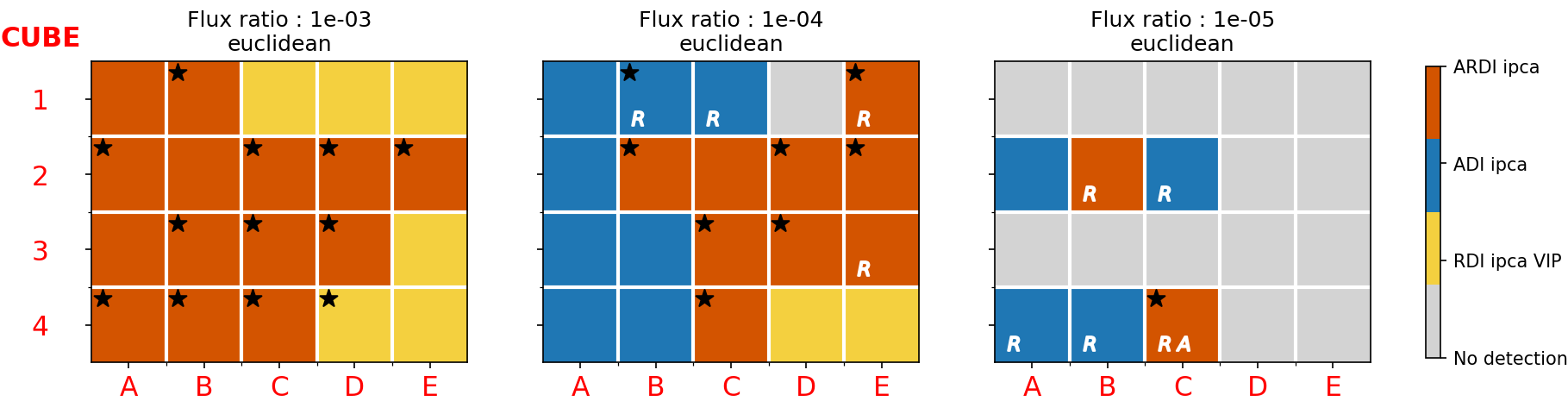}
    \includegraphics[width=0.85\linewidth]{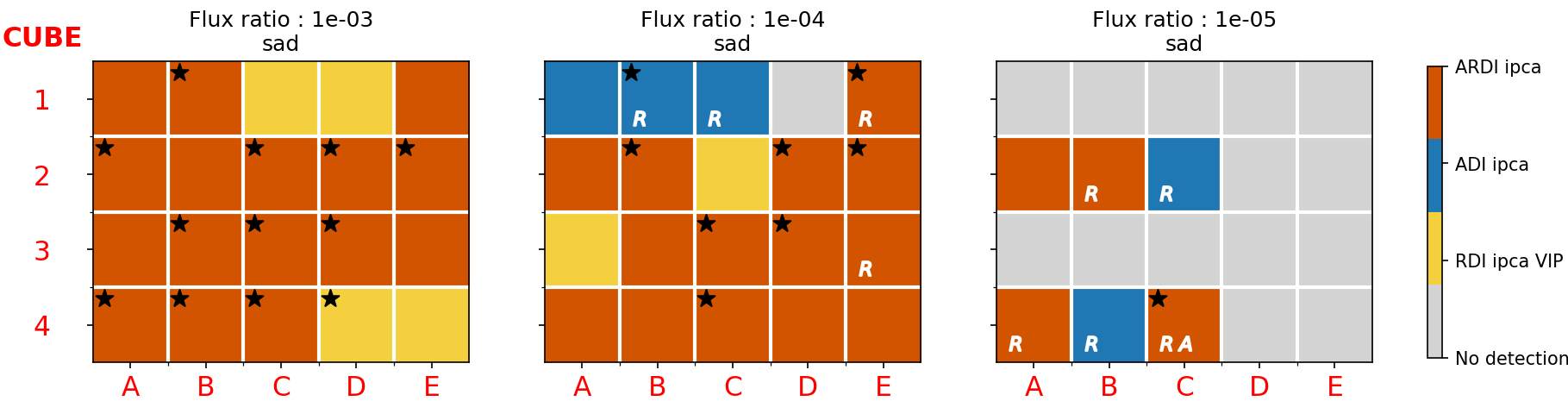}
    \caption{Same as Fig.~\ref{fig:results} but for the four different metrics from top to bottom, respectively: Spearman's rank, Pearson correlation, Euclidean distance, and SAD\remove{ distance}.}
    \label{fig:res_all}
\end{figure*}

This appendix presents the performance of RDI, ADI, and ARDI, using the IPCA algorithm, for each test dataset according to four different metrics (see Fig.~\ref{fig:res_all}: the Spearman rank correlation coefficient, the Pearson correlation coefficient, the Euclidean distance, and the sum of absolute differences (SAD). A table using a fifth metric, the structural similarity index measure \citep[SSIM,][]{SSIM}, was already displayed in Fig.~\ref{fig:results}. In all these figures, the color of the cell indicates which algorithm performed the best according to the respective metrics. White cells mean that no strategy detected the disk. White letters on a cell indicate which strategy did not detect the disk (R$\mapsto$RDI, A$\mapsto$ADI, Ar$\mapsto$ARDI). Black stars indicate that all five metrics (Spearman, SSIM, Euclidean, Spearman, Pearson, and SAD) agree on the strategy which achieved the best disk estimation. Overall, it can be observed that the various metrics yield relatively similar results. ARDI-IPCA is more frequently chosen as the best reconstruction by all metrics, with an average selection rate of 67\% across all metrics (considering detection only). Among the tests where ARDI performed better, 45\% of them are unambiguously better as all metrics agreed. ADI tends to perform better for contrast $10^{-4}$ and for the sharpest disks (A, B) where the deformations due to flux invariant to the rotation are minimal. RDI seems more appropriate for low contrast ($10^{-3}$), stable datasets with well correlated references (\#3) and for the more face-on disks (C, D, E).

\section{Results of disk estimations}
\label{ann:full_results}

This appendix presents the disk estimations obtained with RDI, ADI, and ARDI, using IPCA, for all 60 test datasets considered in this work (Figs.~D.1A to D.4E).\nnew{The figures of this appendix are available on Zenodo: \url{https://zenodo.org/records/11442350}}.

\section{IPCA parameters used to process the images presented in Fig.~\ref{fig:real_disks}}
\label{sec:paramIPCA}

In this appendix, Table~\ref{tab:params} provides a comprehensive summary of the parameters applied in the retrieval of the protoplanetary disk gallery using ARDI-IPCA, as shown in Sect.~\ref{sec:test_real}.
\setlength{\tabcolsep}{2pt}

\begin{table}[!ht]
    \caption{IPCA parameters used to reprocess each dataset from \citet{Ren23a} in Sect.~\ref{sec:test_real}. For each dataset, this table provides the starting rank and the number of iterations per rank. Regarding the number of iterations per rank, ``incr'' stands for incremental and indicates that the number of iterations increases by one at each rank (e.g., one iteration is performed at the starting rank $q$, two at rank $q+1$, and so on). Additionally, we also provide the number of iterations performed to obtain the final frame selected from visual assessment (``final iter.''), \remove{and} the rank that was reached at this iteration (``final rank'')\new{, and the mean and standard deviation of the PCC between science images and references}.}
    \label{tab:params}
\begin{tabular}{l p{0.20\linewidth} p{0.07\linewidth} p{0.07\linewidth} p{0.07\linewidth} 
p{0.07\linewidth} p{0.09\linewidth} p{0.09\linewidth}}
\hline
\hline

Star name   & Date       & Start.\ rank & Iter.\ per rank & Final iter. & Final rank & Mean PCC & STD PCC \\ \hline
CI Tau      & 2021-12-09 & 10           & incr            & 88          & 21         & 0.89     & 0.13    \\
CQ Tau      & 2021-01-01 & 1            & incr            & 85          & 8          & 0.83     & 0.02    \\
CY Tau      & 2021-12-27 & 10           & incr            & 10          & 13         & 0.93     & 0.04    \\
CY Tau      & 2021-12-09 & 10           & incr            & 10          & 13         & 0.94     & 0.02    \\
DL Tau      & 2021-12-04 & 1            & incr            & 10          & 1          & 0.96     & 0.01    \\
DM Tau      & 2020-12-19 & 10           & incr            & 105         & 23         & 0.80     & 0.05    \\
DN Tau      & 2021-11-24 & 5            & incr            & 6           & 7          & 0.79     & 0.25    \\
DN Tau      & 2021-12-10 & 5            & incr            & 6           & 7          & 0.88     & 0.05    \\
DS Tau      & 2021-12-29 & 10           & incr            & 6           & 12         & 0.85     & 0.07    \\
GM Aur      & 2021-01-20 & 5            & incr            & 8           & 9          & 0.80     & 0.04    \\
HD 31648    & 2021-12-10 & 5            & 10              & 17          & 8          & 0.98     & 0.01    \\
V1366 Ori   & 2020-11-28 & 1            & 10              & 41          & 4          & 0.70     & 0.36    \\
V1366 Ori   & 2020-12-24 & 1            & 10              & 46          & 5          & 0.77     & 0.19    \\
V1366 Ori   & 2020-12-27 & 1            & 10              & 41          & 4          & 0.74     & 0.18    \\
HD 97048    & 2021-01-28 & 1            & 10              & 7           & 1          & 0.95     & 0.02    \\
HD 100453   & 2022-06-09 & 1            & 10              & 85          & 8          & 0.45     & 0.06    \\
HD 100546   & 2020-12-22 & 1            & 10              & 6           & 1          & 0.89     & 0.01    \\
HD 143006   & 2021-06-30 & 1            & 10              & 85          & 8          & 0.75     & 0.11    \\
HD 143006   & 2021-07-22 & 1            & 10              & 99          & 9          & 0.83     & 0.10    \\
HD 163296   & 2022-06-11 & 15           & 10              & 145         & 28         & 0.87     & 0.06    \\
HD 163296   & 2021-09-09 & 15           & 10              & 145         & 28         & 0.90     & 0.05    \\
HD 163296   & 2021-09-26 & 15           & 10              & 145         & 28         & 0.92     & 0.03    \\
HD 163296   & 2021-06-03 & 15           & 10              & 145         & 28         & 0.87     & 0.04    \\
HD 163296   & 2021-04-06 & 15           & 10              & 145         & 28         & 0.88     & 0.03    \\
HD 163296   & 2022-07-07 & 15           & 10              & 145         & 28         & 0.92     & 0.05    \\
HD 169142   & 2021-09-06 & 1            & 10              & 31          & 3          & 0.79     & 0.26    \\
IP Tau      & 2021-12-09 & 10           & incr            & 105         & 23         & 0.94     & 0.02    \\
IP Tau      & 2021-12-28 & 10           & incr            & 105         & 23         & 0.91     & 0.03    \\
IQ Tau      & 2022-01-06 & 10           & incr            & 21          & 15         & 0.93     & 0.03    \\
IQ Tau      & 2022-01-01 & 10           & incr            & 91          & 22         & 0.94     & 0.03    \\
IQ Tau      & 2022-01-03 & 10           & incr            & 21          & 15         & 0.95     & 0.02    \\
LkCa 15     & 2020-12-08 & 1            & incr            & 12          & 2          & 0.76     & 0.03    \\
LkCa 15     & 2020-11-27 & 1            & incr            & 85          & 8          & 0.83     & 0.06    \\
LkHA 330    & 2020-12-08 & 1            & incr            & 54          & 10         & 0.89     & 0.01    \\
HD 36112    & 2020-12-23 & 1            & incr            & 22          & 2          & 0.93     & 0.01    \\
HD 36112    & 2020-12-19 & 1            & incr            & 23          & 3          & 0.93     & 0.01    \\
HD 36112    & 2020-12-26 & 1            & incr            & 51          & 5          & 0.92     & 0.01    \\
CPD-68 1894 & 2021-06-04 & 1            & incr            & 85          & 8          & 0.80     & 0.03    \\
V351 Ori    & 2022-02-07 & 1            & incr            & 85          & 8          & 0.78     & 0.02    \\
CPD-36 6759 & 2021-06-04 & 1            & incr            & 85          & 8          & 0.83     & 0.02    \\
SR 20       & 2022-05-12 & 10           & incr            & 22          & 15         & 0.98     & 0.01    \\
SY Cha      & 2021-01-02 & 5            & incr            & 151         & 19         & 0.72     & 0.06    \\
SZ Cha      & 2020-12-29 & 1            & incr            & 85          & 8          & 0.83     & 0.03    \\
SZ Cha      & 2020-12-30 & 1            & incr            & 12          & 2          & 0.83     & 0.03    \\
V1094 Sco   & 2021-09-10 & 10           & incr            & 65          & 20         & 0.80     & 0.05    \\
V1247 Ori   & 2020-12-20 & 1            & incr            & 85          & 8          & 0.75     & 0.04    \\
V1247 Ori   & 2020-12-22 & 1            & incr            & 11          & 1          & 0.75     & 0.06    \\
V1247 Ori   & 2020-12-24 & 1            & incr            & 18          & 2          & 0.72     & 0.06    \\ \hline

\vspace{-10cm}
\end{tabular}
\end{table}
\newpage
\newpage

\section{Detectability of the protoplanet candidates}
\newpage

\new{This appendix presents further analysis for the candidate protoplanets that are not mixed with coincident scattered light signal from the circumstellar disk, thus enabling the possibility of computing a reliable detection limit in terms of contrast, which can then be converted into a mass detection limit. The mass sensitivity is determined assuming no extinction, by converting our contrast limits using ATMO2020 evolutionary models \citep{Phillips20}.%, of detection limits estimated using the new datasets from \citet{Ren23a} in the $Ks$-Band
The contrast curves are computed for a false positive probability equivalent to a $5\sigma$ detection in Gaussian statistics, including the t-student correction for small sample statistics, using routines from the Vortex Image Processing package\footnote{\url{https://github.com/vortex-exoplanet/VIP}} \citep{Gomez17,Christiaens23}. The different subsections of this appendix detail our protoplanet detection limits and provide corresponding contrast curves and translation to masses where relevant.}

\subsection{HD\,36112 (MWC\,758)}
\label{ann:36112}

\begin{figure}[!ht]
    \centering
         \includegraphics[width=\linewidth]{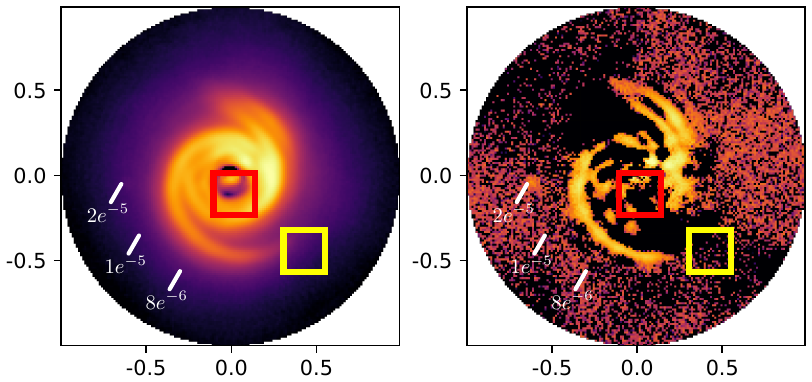}
        \caption{Results of IPCA (left) and 3-rank PCA with ADI only (right) for the VLT/SPHERE dataset of HD 36112 \citep{Ren23a}. Three simulated companions at a separation of $0\farcs6$ have been injected into the cube, with contrasts of $ 8 \times 10^{-6}$, $ 1 \times 10^{-5}$, and $ 2 \times 10^{-5}$. White lines indicate their positions. Two squares are displayed to mark the locations of the two claimed protoplanets: a red square at $0\farcs1$, indicating the candidate proposed by \citet{Reggiani18}, and a yellow square at $0\farcs6$, referring to the candidate proposed by \citet{Wagner19}.}
        \label{fig:36112}
\end{figure}

\begin{figure}[!h]
    \centering
         \includegraphics[width=\linewidth]{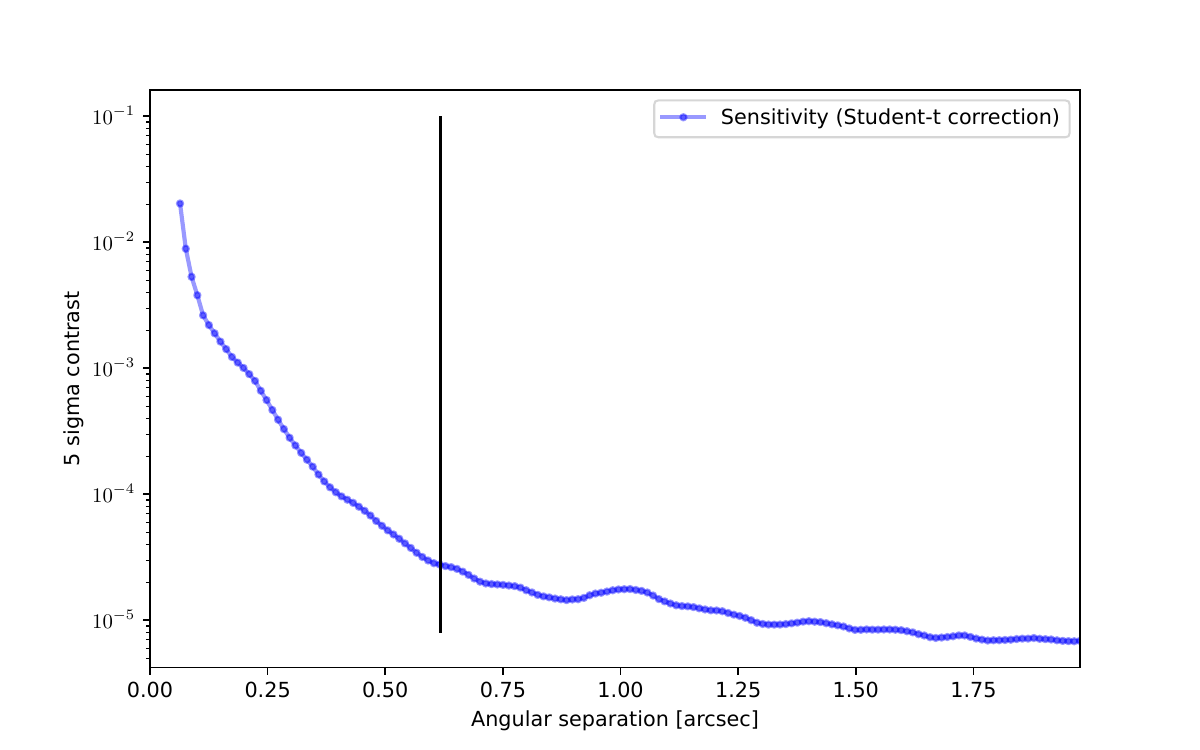}
        \caption{Contrast curve at $5\sigma$ confidence for the HD 36112 dataset, processed using rank-3 PCA with ADI only.}
        \label{fig:cc36112}
\end{figure}

This section presents a test of detectability for the protoplanet candidate proposed by \citet{Wagner19} in the HD\,36112 (MWC 758) system. The test involves injecting fake planets at the same separation as the candidate, using contrasts of $ 8 \times 10^{-6}$, $ 1 \times 10^{-5}$ and $ 2 \times 10^{-5}$. Figure~\ref{fig:36112} shows the results achieved with IPCA (left) on the dataset with the injected planet, in comparison to a result using a rank-3 PCA based on ADI only (right). Regarding the chosen rank for PCA, we processed the data for ranks 1 to 10 and identified the most effective \remove{estimate}\new{rank} for retrieving the fake companion through visual assessment. Additionally, in Fig.~\ref{fig:cc36112}, we present a $5\sigma$ contrast curve computed with rank-3 PCA using ADI only. We observe that only the simulated candidate at $ 2 \times 10^{-5}$ is successfully recovered. The injected simulated planet is distinctly visible after post-processing the data with PCA and ADI-only, but this approach also introduces numerous point-like features attributable to the filtered disk. However, \new{this same injected companion at $ 2 \times 10^{-5}$} is \remove{indistinguishable}\new{undetectable} \remove{with IPCA using ARDI with}\new{in the image recovered with IPCA+ARDI considering} the chosen rank and iteration. This is because the feature becomes blended with the brighter disk halo when the entire disk is recovered. The  $5\sigma$ contrast curve with the rank-3 PCA corroborates our observation and shows a sensitivity reaching a contrast of $\sim 2.6 \times 10^{-5}$ at the separation of the candidate ($0\farcs6$)\new{, which is consistent with the $3 \times 10^{-5}$ contrast limit established by \citet{Grady13} at this separation.} \new{According to the ATMO2020 model and assuming no extinction, our contrast threshold translates into a $1M_{\text{Jup}}$ sensitivity. This mass sensitivity does not match with the predicted mass inferred by \citet{Wagner19}, which was $2-3M_{\text{Jup}}$, using the COND ``hot-start'' model \citep{Baraffe03, Baraffe15}. Indeed, the ATMO2020 model primarily focuses on the atmospheric properties of exoplanets, while the COND hot-start model focuses on the early stages of giant planet formation, particularly the rapid accretion of gas and dust from a protoplanetary disk and the subsequent generation of internal heat.}
\new{Nevertheless, to enable a consistent comparison, we also convert the contrast of $1 \times 10^{-5}$ for the candidate in the $L'$ band as presented in \citet{Wagner19} into a mass using the ATMO2020. We found a mass of $0.5 M_{\text{Jup}}$, suggesting that our detection limit of $1M_{\text{Jup}}$ would not allow us to detect the candidate.} 
\new{All quoted mass conversions assume negligible extinction from circumplanetary and circumstellar material, which is likely optimistic. Inclusion of extinction would inflate our mass limits, affecting more the mass inferred from the Ks-band contrast than the mass inferred from the L' band.}
%However, a contrast inferior to $ 10^{-5}$ is expected in the $K$ band for the candidate companion of \citet{Wagner19}, which is beyond the detection limit inferred in the present tests.

Regarding the protoplanet proposed by \citet{Reggiani18}, we can observe in the images that complex structures arise at the separation of the candidate, including point-like features, especially when using PCA with ADI-only. This could be explained by filtered disk signal.

\subsection{HD\,169142}
\label{ann:169142}

\begin{figure}[!ht]
    \centering
         \includegraphics[width=\linewidth]{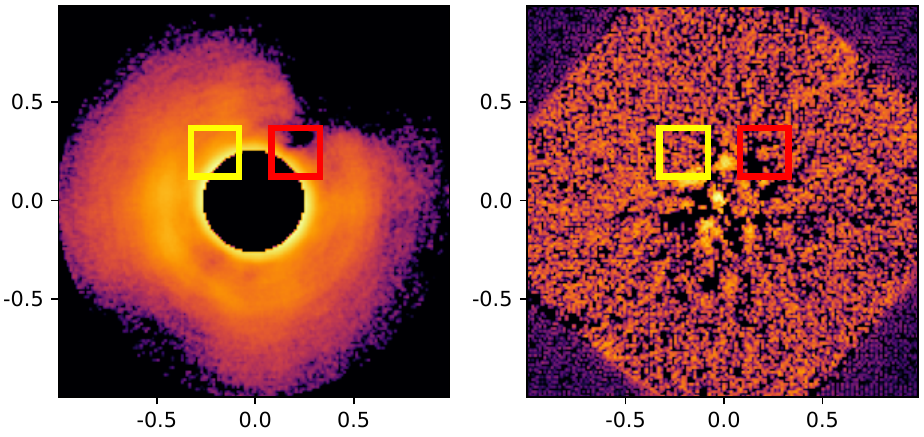}
        \caption{Results of IPCA with ARDI (left) and rank-3 PCA with ADI only (right), for an injected companion at a separation of $0\farcs31$ into the VLT/SPHERE dataset of HD 169142 \citep{Ren23a}, using a contrast of $ 1.5 \times 10^{-5}$. The position of the simulated candidate is indicated by a red square, while the location of the actual candidate proposed by \citet{Gratton19, Hammond23} is marked by a yellow square}
        \label{fig:169142}
\end{figure}

\begin{figure}[!ht]
    \centering
         \includegraphics[width=\linewidth]{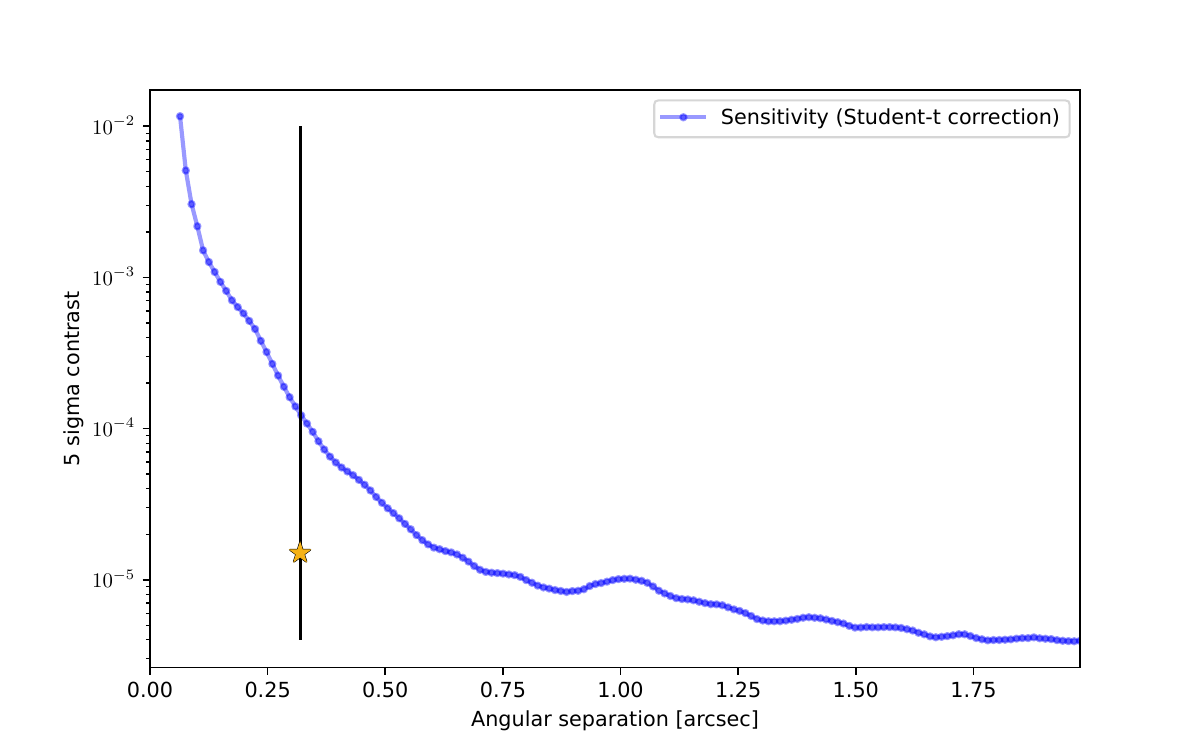}
        \caption{Contrast curve at $5\sigma$ confidence for the HD 169142 dataset, processed using PCA with ADI only. The separation of the candidate at $0\farcs31$ is indicated by a black vertical line. The yellow star indicate the expected contrast of the candidate, inferred by \citet{Hammond23}.}
        \label{fig:cc169142}
\end{figure}

This section presents a test of detectability for the protoplanet candidate proposed by \citet{Gratton19} and \citet{Hammond23} in the HD\,169142 system. The test involves injecting a fake planet candidate at the same separation of $0\farcs31$, using a contrast of $ 1.5 \times 10^{-5}$ as estimated by \citet{Hammond23}. Similarly as in Appendix~\ref{ann:36112}, Fig.~\ref{fig:169142} shows the results of IPCA and rank-3 PCA post-processing with ADI-only results for the dataset with the injected planet. Figure~\ref{fig:cc169142} presents the $5\sigma$ contrast curve using rank-3 PCA with ADI-only. In Fig.~\ref{fig:169142}, the injected planet is barely visible, and only for the rank-3 PCA processing. Although it is at the noise level, it is still discernible due to its clear surrounding, likely a result of the over-subtraction effect. According to the contrast curve in Fig.~\ref{fig:cc169142}, the sensitivity at the separation of $0\farcs31$, where the candidate is located, is $ 1 \times 10^{-4}$.
%\new{We notice a significant disparity between the contrast limit and the detection via injected planet, which might be related to the t-stutend correction, is not comparable to Gaussian $5\sigma$ SNR.}
%\remove{We noticed a disparity between the contrast limit inferred via the contrast curve and the detection via injected planets, which might be related to the T-student correction. The $5\sigma$ SNR for which the T-student correction is applied is not entirely comparable to the Gaussian $5\sigma$ SNR.} 
Moreover, PCA also reveals numerous point-like features due to the filtered bright inner disk. No such PLF appears at the location of the actual protoplanet candidate. % However, the marginal detection of the simulated planet raises uncertainty about the observability of the protoplanet candidate, given the flux level and separation suggested by \citet{Hammond23}. 
\new{Our detection limit of $1 \times 10^{-4}$ translates to $2 M_{\text{Jup}}$, according to the ATMO2020 model, assuming no extinction. In comparison, the contrast in $YJH$ bands reported by \citet{Hammond23} corresponds to $1 M_{\text{Jup}}$. Nevertheless, these results must be put into perspective regarding the presence of signal at the location of the protoplanet in polarized intensity images shown in \citet{Hammond23}, suggesting the presence of circumplanetary dust. Hence, the usage of the ATMO2020 model and the assumption of no extinction might not be appropriate.}
Therefore, the sensitivity limits in our data do not allow us to confirm or refute the candidate.

\subsection{HD\,97048}
\label{ann:97048}

\begin{figure}[!ht]
    \centering
         \includegraphics[width=\linewidth]{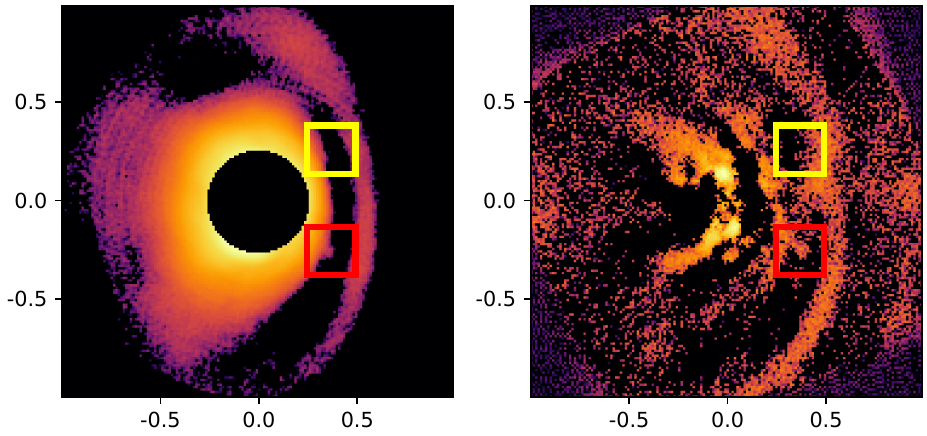}
        \caption{Results of IPCA (left), and rank-3 annular-PCA (right) with ADI only, for a fake companion at a separation of $0\farcs45$, using a contrast of $ 3 \times 10^{-5}$ injected into the VLT/SPHERE dataset of HD 97048 \citep{Ren23a}. The location of the fake candidate is indicated by a red square, while the location of the actual protoplanet candidate inferred through the observation of kinks by \citet{Pinte18} is indicated with a yellow square.}
        \label{fig:97048}
\end{figure}

This section presents a test to infer the detection limit in the $Ks$-band for the protoplanet candidate deduced by \citet{KINKHD97048} using local deviations in the Keplerian velocity in the disk surrounding HD\,97048. The test involves injecting several fake planet candidates at the same separation as the candidate. For an injected planets at a contrast of $3 \times 10^{-5}$, we observe a marginal detection. Fake planets injected with a contrast below $ 1 \times 10^{-5}$ are not recovered, indicating that we cannot detect planets smaller than one \remove{few} Jupiter mass\remove{es} with our images, according to the ATMO2020 model, assuming no extinction. Figure~\ref{fig:97048} displays the image with the fake planet at contrast $ 3 \times 10^{-5}$, obtained with IPCA (left) and with 3-rank PCA with ADI-only (right). The injected simulated planet is visible with both PCA with ADI-only and IPCA with ARDI results. The PCA estimation also reveals numerous point-like features due to the filtered bright inner disk. Notably, no such point-like feature appears at the location of the actual protoplanet candidate. No contrast curve is provided for this dataset, as there is a disk signal at the same separation as the candidate.

\begin{figure*}[!t]
    \centering
         \includegraphics[width=\linewidth]{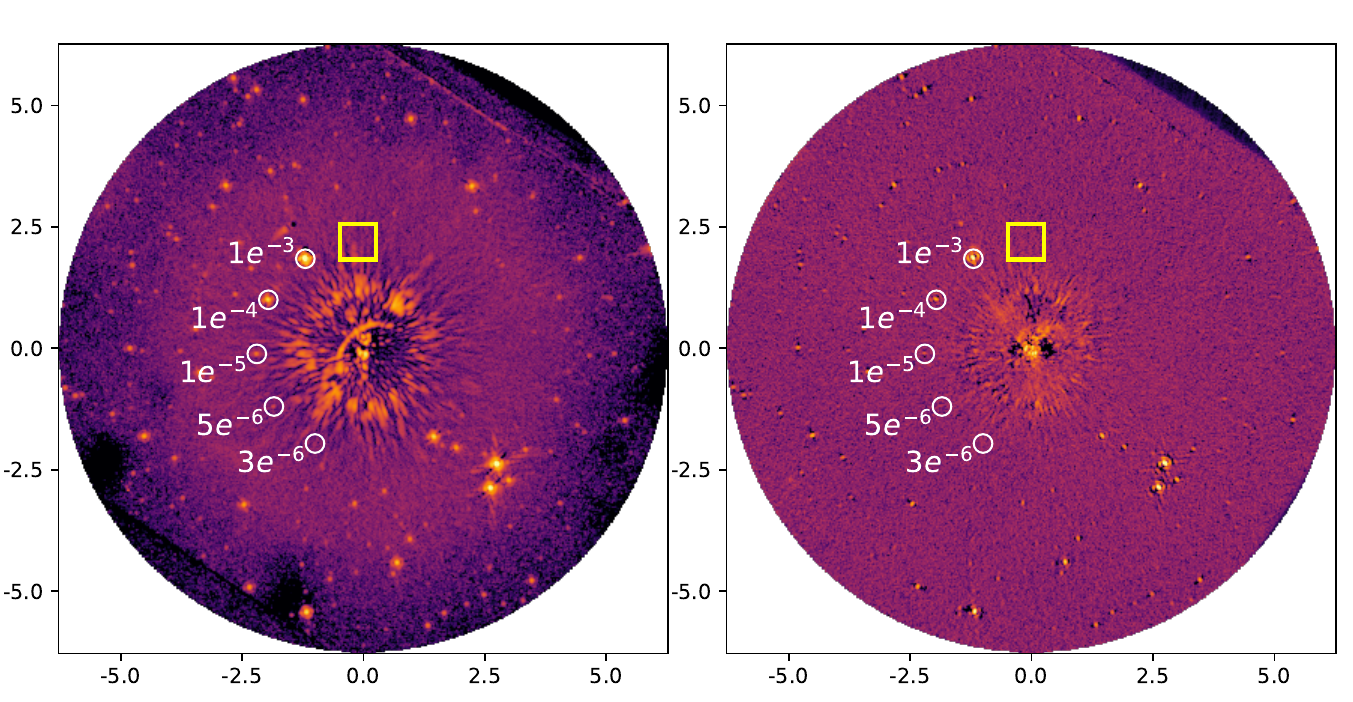}
        \caption{Results of IPCA (left) and rank-2 PCA  (right) both with ADI only, for the uncropped observations of HD 163296. The position of the candidate, inferred by \citet{163296KINK} and \citet{Pinte18} from a local deviation in Keplerian velocity, is marked by the yellow square with dimensions $60 \times 60$ pixels. A zoom into the $60 \times 60$ pixels aperture is displayed in the bottom left of each image, with the planet's location indicated by an arrow on the smaller images. Five simulated companions at a separation of $2\farcs2$ have been injected into the cube, with contrasts ranging from $1 \times 10^{-3}$ to $3 \times 10^{-6}$. White circle indicate their positions.}
        \label{fig:163296}
\end{figure*}

\subsection{HD\,163296}
\label{ann:163296}

\begin{figure}[!t]
    \centering
         \includegraphics[width=\linewidth]{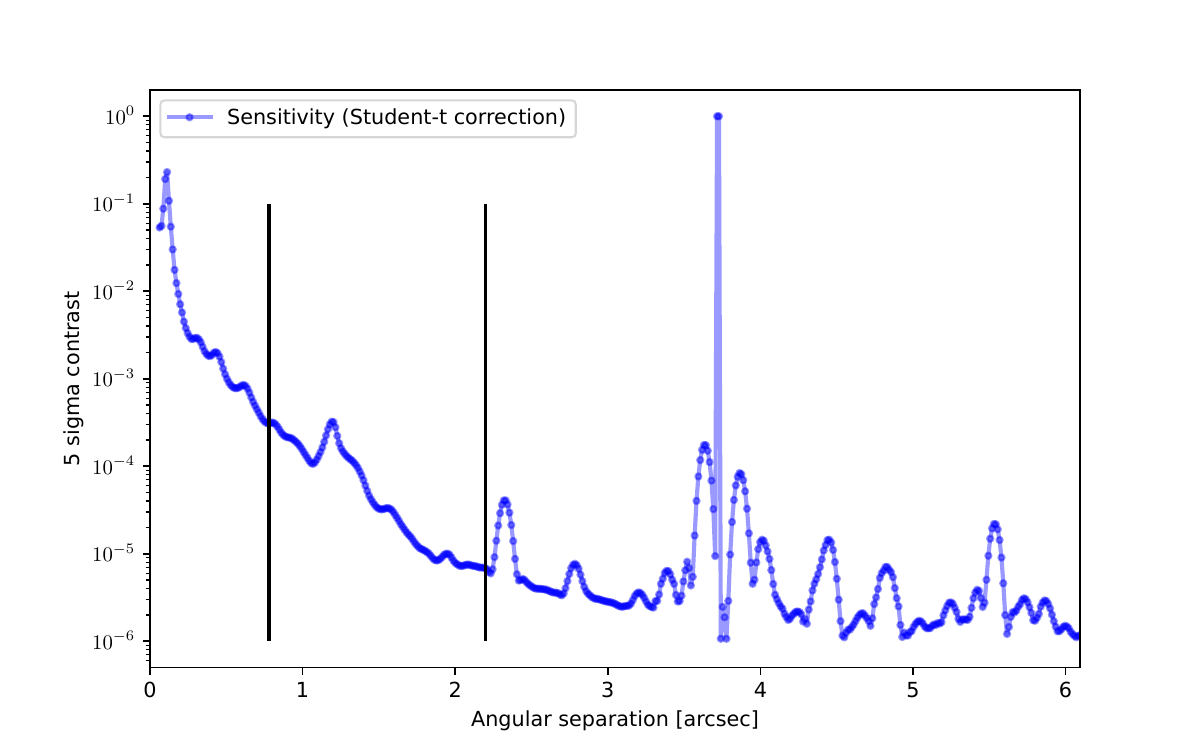}
        \caption{Contrast curve at $5\sigma$ confidence for the HD 163296 dataset, processed using PCA with ADI only. The separation of the candidate, inferred by \citet{163296KINK} and \citet{Pinte18} from a local deviation in Keplerian velocity at $2\farcs2$, is indicated by a black vertical line.}
        \label{fig:contrastcurve}
\end{figure}

This section presents the $6\arcsec$-width uncropped observations of HD\,163296 to check for any possible signal arising at the location of the detected kinks at 260\,au ($2\farcs2$) from the star \citep{163296KINK, Pinte18}. We present in Fig.~\ref{fig:163296} the results of the processed data using IPCA and PCA, both with ADI-only, and contrast curve is shown in Fig.~\ref{fig:contrastcurve}, indicating an achieved $5\sigma$ contrast of $6\times 10^{-6}$ at a separation of $2\farcs2$. In Fig.~\ref{fig:163296}, no signal is observed at the candidate location, suggesting that the companion might have a contrast below the detection limit of $ 6\times 10^{-6}$. We injected multiple simulated companions at a separation of $2\farcs2$, with contrasts ranging from $1 \times 10^{-3}$ to $3 \times 10^{-6}$. We found that for a contrast below $6\times 10^{-6}$, the injected companion is hardly differentiable from remaining speckle patterns that have not been well subtracted. \remove{a contrast of }Additionally, the IPCA results reveal a halo spanning approximately from $2\farcs3$ to $4\farcs5$. In a previous study using data from the \nnew{\textit{Hubble}} Space Telescope, \citet{Grady00} reported extended signal between $2\farcs9$ and $3\farcs2$ in the southeast and northeast of HD\,163296. However, our observation exhibits a distinctively circular halo, which does not match the disk inclination, suggesting that this signal may not originate from a circumstellar source and may not be related to the signal observed by \citet{Grady00}.

For the kink observed at a separation of $0\farcs85$, our detection limit at 5$\sigma$-contrast of $3 \times 10^{-4}$, translates to an apparent magnitude of $M_{Ks} \sim 13.6$. According to the ATMO2020 evolutionary models \citep{Phillips20}, this magnitude corresponds to a mass sensitivity of approximately \remove{10}\new{4} Jupiter masses, while neglecting extinction. Hence, our observations may not reach sufficient depth to detect a $Ks$-band counterpart for this particular kink.

For the kink situated at $2\farcs2$, our detection limit at 5$\sigma$-contrast of $8 \times 10^{-6}$ corresponds to an apparent magnitude of $M_{Ks} \sim 17.8$. This sensitivity level is lower than 1$M_{\text{Jup}}$ (with $M_{Ks} <$ 16.3) based on the ATMO2020 evolutionary models and approximately 1 Jupiter mass ($M_J \sim$ 17.5) according to the BEX models \citep{Linder19}, assuming no extinction effects. Notably, based on the amplitude of the kink, \citet{Pinte18} suggested a potential 2$M_{\text{Jup}}$ planet, which would imply an expected apparent magnitude of $M_{Ks} \sim$ 15.6 based on the BEX models. This discrepancy suggests a significant extinction effect in the $Ks$ band, estimated to be $A_{Ks} \gtrsim 2.0$.

\end{appendix}

\end{document}